\documentclass[12pt,preprint]{aastex}

\newcommand{\et}{et al.}
\newcommand{\kms}{km s$^{-1}$}
\newcommand{\ha}{H$\alpha$}
\newcommand{\solar}{\ifmmode_{\sun}\;\else$_{\sun}\;$\fi}

\newcommand{\logsfrd}{$\log \dot{M}_{D}$}
\newcommand{\HII}{H$\,${\sc ii}}
\newcommand{\HI}{H$\,${\sc i}}

\begin{document}

\title{Mid-Infrared Images of Stars and Dust in Irregular Galaxies\footnotemark[1]}

\footnotetext[1]{This work is based in part on archival data obtained with the 
Spitzer Space Telescope, which is operated by the Jet Propulsion Laboratory, 
California Institute of Technology under a contract with NASA.}

\author{Deidre A. Hunter\altaffilmark{2},
Bruce G. Elmegreen\altaffilmark{3},
and Emily Martin\altaffilmark{2,4}}

\altaffiltext{2}{Lowell Observatory, 1400 West Mars Hill Road, Flagstaff,
Arizona 86001 USA}

\altaffiltext{3}{IBM T. J. Watson Research Center, PO Box 218, Yorktown 
Heights, New York 10598 USA}

\altaffiltext{4}{Current address Wheaton College, Norton MA 02766}

\begin{abstract}
We present mid-infrared to optical properties
of 22 representative irregular galaxies: 18 Im, 3 BCDs, and one Sm.
The mid-IR is based on images 
from the {\it Spitzer Space Telescope} archives. 
The 3.6 and 4.5 $\mu$m bands and the $UBVJHK$ images
are used to examine disk morphology 
and the integrated and 
azimuthally averaged magnitudes and colors of stars.
The non-stellar contribution to the 4.5 $\mu$m images is used
to trace hot dust.
The 5.8 and 8.0 $\mu$m images reveal
emission from hot dust and PAHs, and both may contribute to these passbands,
although we refer to the non-stellar
emission as PAH emission.
We compare the 8.0 $\mu$m images to \ha.
Im galaxies have no hidden bars,
and those with double-exponential 
optical light profiles have the same at mid-IR.
Most galaxies have similar optical and mid-IR scale lengths.
Four galaxies 
have super star clusters that are not visible at optical bands.
Galaxies with 
higher area-normalized star formation rates have more dust and PAH emission
relative to starlight.
Hot dust and PAH emission
is found mostly in
high surface brightness \HII\ regions, 
implying that massive
stars are the primary source of heating. 
Galaxies
with intense, wide-spread star formation have more extended
PAH emission. 
The ratio of PAH to \ha\ emission
is not constant on small scales.
PAHs are associated with
shells and giant filaments, so
they are not destroyed during shell formation.
\end{abstract}


\section{Introduction} \label{sec-intro}

The {\it Spitzer Space Telescope} ({\it Spitzer}) allows us
a deeper view into the infrared emission of galaxies
than was possible with previous infrared satellites.
The InfraRed Array Camera (IRAC;
Fazio \et\ 2004) in particular gives us the opportunity
to image galaxies in passbands that capture the light
of stars and emission from hot dust and 
polycyclic aromatic hydrocarbons (PAHs) 
with a spatial resolution that
is comparable to that of ground-based optical images.
IRAC's Channel 1 (3.6 $\mu$m) centered at 3.6 $\mu$m probes the stellar population;
Channel 2 (4.5 $\mu$m) centered
at 4.5 $\mu$m contains emission from hot dust as well as stars;
and Channel 3 (5.8 $\mu$m) at 5.8 $\mu$m and Channel 4 (8.0 $\mu$m) at 7.8 $\mu$m are
dominated by thermal emission from PAHs in the interstellar medium (ISM)
along with emission from hot dust.
The 8.0 $\mu$m image in particular contains the 7.7 and 8.6 $\mu$m emission
bands due to PAHs, and the 5.8 $\mu$m image contains the 6.2 $\mu$m band
(L\'eger \& Puget 1984, Allamandola \et\ 1989).

Irregular (Im) galaxies, the most numerous type of galaxy
in the universe, have received limited attention with infrared
satellite telescopes because of their faintness which makes
observations difficult. However, they are an interesting class
to examine in the infrared because of their distinct differences compared
to giant spiral systems. As a class the
Im galaxies have a lower dust content, and their molecular clouds are,
to a greater extent than spirals, the cores of atomic clouds.
They also are generally dominated in the optical
by their young stellar populations. They are 
physically smaller, lower in luminosity and
surface brightness, and more gas-rich.
Most Im galaxies are forming
stars; yet they do so without the benefit of spiral density
waves. The Im galaxies may represent the nature of star formation in the
early universe.

Generally, PAH emission is low in dwarf galaxies, and this
may be related to the low metallicity and intense stellar radiation field
in these systems relative
to those of spirals (Boselli \et\ 1998, Thuan \et\ 1999,
Madden 2000, Sturm \et\ 2000,
Galliano \et\ 2003, Lu \et\ 2003, Houck \et\ 2004, Rosenberg \et\ 2006).
Hunter \et\ (2005), for example, used {\it Infrared Space Observatory} mid-infrared
imaging (6.75 $\mu$m and 15 $\mu$m)
and far-infrared (FIR) spectroscopy 
to examine the properties of five Im systems. They
found that 
the PAH emission, which is associated only with the brightest \HII\ regions,
is depressed
relative to that of small grains, the FIR, and \ha.
In addition, the integrated [CII] emission relative to PAH
emission is unusually high compared to that in spirals.
These two observations suggest that Carbon is more likely to be in 
an atomic form than a PAH.
Recently, in a study of Sloan Digital Sky Survey 
galaxies observed in the IRAC First Look Survey,
Hogg \et\ (2005) found that low luminosity blue galaxies have
anomalously low
PAH emission relative to starlight.
Similarly, Engelbracht \et\ (2005) have examined a sample of low metallicity systems,
and found that the 8.0 $\mu$m-band nebular emission decreases abruptly
relative to the 24 $\mu$m emission from dust for galaxies with metallicities
less than 1/3--1/5 solar.
As Draine (2005) has pointed out, the ISM of low metallicity
systems seems to be a more
hostile environment for PAHs for reasons that are not entirely clear.

Here we use the unique capabilities of the IRAC camera to
further our understanding of the stars and dust in dwarf galaxies through 
an investigation of 22 irregular galaxies in the mid-IR.
We examine the distribution and intensity
of hot dust and PAH emission in detail and compare 
them to that of \ha, a tracer of
the current star formation activity.
We also probe the stellar populations with the 3.6 $\mu$m and 4.5 $\mu$m images,
determining the stellar distributions and colors and comparing
the mid-IR to the optical.

The sample includes 18 Im systems, 3 blue compact dwarfs (BCDs),
and one Magellanic-type spiral (Sm) galaxy. 
Except for NGC 4449, these are systems included in the 
\ha\ and broad-band survey of Im/BCD/Sm galaxies conducted
by Hunter \& Elmegreen (2004, 2006).
The galaxies and some pertinent parameters are listed in
Table \ref{tab-sample}. 
The galaxies included in this study span a large range in
properties: $M_V$ ranges from $-$11.7 to $-$18.3; the star
formation rate ranges from no detectable \ha\ emission
(M81dwA and DDO 210) to galaxies undergoing a burst of
star formation (IC 10, NGC 1705). This sample, 
although defined by what was available in the {\it Spitzer} archives,
is fairly representative of the Im class and is not
dominated by high star formation rate dwarf galaxies which have 
been the subject of most past studies of dwarfs in the infrared. 
Furthermore, except for NGC 4449, these are relatively isolated systems;
the galaxies are not involved in an obvious on-going interaction
with another galaxy. 

\section{Data Reduction and Analysis} \label{sec-data}

The galaxies were imaged with the IRAC detectors; 
16 of the galaxies were imaged in all four pass-bands
and 6 were imaged only at 4.5 $\mu$m and 8.0 $\mu$m. The imaging
employed various dithering patterns, and large galaxies
were mapped. All galaxies had short and long or
short, medium, and long exposures. Many of the galaxies
were observed in two separate observing sessions
(those from the SINGS Legacy database, Kennicutt \et\ 2003).

The data were obtained from the {\it Spitzer} archive and 
reduced with the Basic Calibrated Data pipeline.
Post-pipeline reductions were done using the {\it Spitzer}
contributed software
package MOPEX. Saturated pixels in the long exposures were
replaced with appropriate values determined from the short or
medium exposures. Overlapping fields were placed onto a single
grid determined to include all frames for each galaxy, and backgrounds
were matched. 
Optical distortion was removed and the images were repixellated
to 1.2\arcsec\ per pixel.
We used dual-outlier and mosaic-outlier detection
algorithms to remove cosmic rays, and experimented with different 
outlier rejection parameters before settling on thresholds
and other parameters that best removed the cosmic rays and
bad pixels without removing real objects.
We did not do corrections for ``muxbleed'' or ``column
pull-down'' problems since the corrections are not adequate for photometry;
we preferred to be able to identify these flaws in the
final images so that they could be masked and excluded in
the photometry.

We determined the background 
in the images themselves from regions not containing galaxy. In some cases
the background was sufficiently flat that a simple constant was
used. In other cases, we did a two-dimensional fit, usually
of order two, to the background. The background was subtracted 
from the image.

To prepare the images for photometry, we needed to remove the
foreground stars and background galaxies contaminating the 
object galaxy image, as we had done for the UBVJHK images
(Hunter \& Elmegreen 2006).
We used the $V$-band
image of the galaxy and surrounding non-galaxy regions in the IRAC
images as guides
to what was likely to be galactic. 
In most cases we edited objects
from the image by interpolating across a circle centered on the star
to be removed.
The editing was done star by star for the 3.6 $\mu$m image, and
the commands were logged to a file;
then the logged commands were applied to the 4.5 $\mu$m image so that 
the 3.6 $\mu$m and 4.5 $\mu$m images
were cleaned in an identical manner. 

In some cases this method of removing foreground stars was not
feasible (DDO 69, DDO 75, DDO 216, IC 10, NGC 3109, NGC 4214,
and NGC 6822) either because the galaxy itself was highly resolved
or the foreground contamination was particularly high. In those cases
we used the region surrounding the galaxy to determine a constant
background that included a statistical sampling of foreground stars.
We tried both techniques on a few galaxies, particularly
those in which the slopes in
the exponential disks were different in 3.6 $\mu$m than in $V$. We found
that the choice of technique for foreground subtraction
did not make any significant difference to the photometry.

In NGC 6822, Milky Way cirrus was prevalant in the 8.0 $\mu$m image.
Thus, we did not extract quantitative information, but we did
make a qualitative spatial comparison with the \ha\ image,
restricting ourselves to the higher surface brightness emission.

The IRAC images were flux calibrated and pixel values converted to units
of MJy sr$^{-1}$ in the Basic Calibrated Data pipeline. 
We express quantities here as magnitudes
relative to Vega where
we have used fluxes of 280.9, 179.7, 115.0, and 64.1 Jy for Vega
for bands 3.6 $\mu$m, 4.5 $\mu$m, 5.8 $\mu$m, and 8.0 $\mu$m, respectively
(Reach \et\ 2005),
and we will refer to these images and quantities measured from them
as [3.6], [4.5], [5.8], and [8.0], respectively.
The Johnson $UBV$ photometric system is also normalized to Vega, so
the IR-optical combination of $V-$[3.6] that we will use is in terms
of the Vega IR-optical color.
The calibration is based on point sources with photometry performed
through a 12\arcsec\ radius aperture. An aperture correction
to extended sources is, therefore, necessary.
The corrections
we adopted are 0.063, 0.071, 0.281, and 0.331 mag for
[3.6], [4.5], [5.8], and [8.0], respectively, from Reach \et\ (2005).

No reddening corrections have been applied to the IRAC 
photometry. Internal reddening in irregular galaxies is low
(Hunter \et\ 1986, 1989),
and in most cases, the foreground reddening is zero or near zero.

The [4.5] passband contains emission from stars and hot dust, and
the [5.8] and [8.0] passbands include PAH emission in addition to
starlight. 
We will refer to the hot dust
emission in [4.5] as ``[4.5]$_{\rm dust}$'' and that from the stars
as ``[4.5]$_*$.''
The [5.8] and [8.0] images also contain emission
from warm dust (Rosenberg \et\ 2006). PAH emission-bands are
found in the spectra of Im galaxies,
so emission from both PAHs and dust could,
in principle be present (Engelbracht \et\ 2005). 
We will usually refer to 
the non-stellar emission in [8.0] as ``PAH emission'' 
for simplicity in order to distinguish it from stellar emission, but
keep in mind that both PAHs and hot dust may contribute to the
[5.8] and [8.0] passbands. All quantities measured from
[5.8] and [8.0] are measured from the star-subtracted images,
and are referred to as ``[5.8]$_{\rm PAH}$'' and
``[8.0]$_{\rm PAH}$''.

We have used the [3.6] images to remove the
stars from the other passbands and leave non-stellar emission.
To do this, we used foreground stars in each image to determine
a scaling factor that matches the stellar photometry in the [3.6] 
image to that in
the other images, scaled the [3.6] image,
and subtracted the scaled [3.6] image. The scaling was done empirically,
so no explicit assumption
was made concerning the spectral type of the stars.
In the case of [4.5], this sometimes resulted in an oversubtraction
of the galaxy (IC 10, NGC 3738, NGC 4449), 
presumably because the galaxy is blue and the foreground
stars are red. In those cases we empirically adjusted the scaling factor 
by 5--11\% to make the stars go away without oversubtracting the
galaxy itself.
The average factor that the 3.6 $\mu$m images were divided by were
$1.59\pm0.08$, $2.44\pm0.18$, and $4.23\pm0.30$ for the 4.5 $\mu$m,
5.8 $\mu$m, and 8.0 $\mu$m images, respectively. The uncertainty
is the variation between galaxy fields.
For those galaxies with only 4.5 $\mu$m and 8.0 $\mu$m images, the
4.5 $\mu$m image was divided by $2.72\pm0.14$, on average, before 
being used to remove stars from the 8.0 $\mu$m image.
For those galaxies as well, a [4.5]$_{\rm dust}$ image could not be
constructed and the original [4.5] image was taken to be [4.5]$_*$.

Having constructed a hot dust image as just described, we subtracted
the hot dust from the original [4.5] image to leave an image of
only starlight in this passband. According to Regan \et\ (2004)
we should expect most of the emission in the [4.5] image to be
starlight for galaxies undergoing quiescent star formation,
and indeed, the contribution from hot dust was small in our
systems.

In the highly resolved galaxy NGC 6822 we found that there 
were stars left in the [4.5]$_{\rm dust}$ and [5.8]$_{\rm PAH}$ images after scaling
and subtracting the [3.6] image. These stars clustered around
NGC 6822, so they are likely stars in NGC 6822 rather than
foreground objects. We were curious as to the nature of these stars.
Therefore, we measured
brightnesses of those stars as well as of a sample
of NGC 6822 stars that did subtract cleanly. We did
the photometry on the IRAC images as well as on $JHK$
2MASS images that we extracted from the 2MASS archive.
We compared the location of the stars on the J$-$H 
vs H$-$K color-color diagram with stellar photometry
of the LMC presented by Nikolaev \& Weinberg (2000).
Stars that had subtracted cleanly lie in the region of
the color-color diagram populated by dwarf and giant
stars. Stars that did not subtract successfully
lie where one expects Carbon stars and B[e]
stars to sit. 
Thus, it is likely that the stars that remained in the [4.5]$_{\rm dust}$ and
[5.8]$_{\rm PAH}$ images
after scaling and subtracting the [3.6] image are real objects in
NGC 6822. To construct the non-stellar dust and PAH images,
these stars were edited out of the [4.5]$_{\rm dust}$ and [5.8]$_{\rm PAH}$ images by hand.

For most of our sample galaxies, we have \ha\ and $UBV$ images,
and for a few we also have $JHK$ images (Hunter \& Elmegreen 2004, 2006).
In order to compare the galaxies' emission in these different passbands,
we geometrically transformed all images to match the scale
and orientation of the $V$-band image.
Surface photometry was extracted in ellipses using the 
parameters---center, ellipticity, major axis step size---that were used
for photometry of the $V$-band image.
The photometry parameters for all galaxies except
NGC 4449 are listed by Hunter \& Elmegreen (2006).
The original optical surface photometry for NGC 4449 was presented
by Hunter \et\ (1999) who used a position angle and ellipticity
that changed with radius
because of the substantial stellar bar.
Here we have used the position angle (64\arcdeg)
and ellipticity ($b/a=0.73$) that are appropriate
for the outer disk to be consistent with our other photometry, 
and have redone the optical surface photometry
to match.

In the ellipse
surface photometry, bad pixels were masked and replaced
with an average of the brightness in the annulus. Some galaxies
were not completely covered on one or more sides by the IRAC maps
(IC 10, NGC 4214, NGC 4449). The portion of the galaxy that was
blank was treated as for masked pixels, and the surface photometry
was truncated when the portion that was missing became larger than
the portion with measurable galaxy. The IRAC image of
NGC 6822 has a field of view that is large enough to include the
entire galaxy, whereas our optical image does not.
As a result, the IRAC surface photometry only goes
as far out in radius as the optical goes, but for integrated IRAC values 
we used the original IRAC image and its larger field of view.

The uncertainties in the surface photometry followed the formula
given by Hunter \& Elmegreen (2006). The original sky background
was reconstructed for this purpose using the frame duration time and
estimates of the zodiacal and Galactic backgrounds
(FRAMTIME, ZODY\_EST,
and ISM\_EST key words in the header of the images).

The galaxies are shown in Figure \ref{fig-pix}. There are potentially
3 images, all showing the same field of view, and 
two graphs for each galaxy. These are arranged in two rows per galaxy
and two galaxies per page. 
The upper left image for each galaxy is the geometrically transformed
[3.6] image, or logarithm of [3.6] if the 
galaxy is high in surface brightness. 
An outer $V$-band contour is plotted on this
image for comparison. For the galaxies with no [3.6] image, the
[4.5] image is shown instead. The upper middle image is an \ha\ 
image with a contour of the [8.0]$_{\rm PAH}$ PAH emission image superposed.
The lower left image is the same but contours are of the 
[4.5]$_{\rm dust}$ image. If the galaxy has no detectable hot dust, this image 
is left blank. If the galaxy has no detectable PAH emission,
the \ha\ image is shown with the same outer $V$-band contour
superposed that was used on the [3.6] or [4.5] image.
We do not show \ha\ images for M81dwA and DDO 210 because
no \ha\ emission was detected in these galaxies. In those
cases there is also no hot dust or [8.0]$_{\rm PAH}$ PAH emission,
so nothing is shown in the upper middle and bottom left spots.
The graph in the bottom middle spot is the surface photometry
in $V$, $J$ if observed, \ha, [3.6] or [4.5]$_*$, and [8.0]$_{\rm PAH}$ PAH emission,
if detected, as a function of radius. The solid lines are
fits to the exponential disks in $V$, $J$, and [3.6] or [4.5]$_*$.
The graph on the right is a plot of the colors as a function
of radius. 

Integrated photometry 
is given in Table \ref{tab-int} along with the
central surface brightness $\mu_0^{[3.6]}$ and
disk scale length $R_D^{[3.6]}$ determined from the exponential fit
to the surface photometry
for the galaxies with observations in all 4 IRAC
passbands. Similar quantities are given in Table \ref{tab-int2}
for the galaxies with observations in only [4.5] and [8.0] 
passbands. Upper limits are 5$\times$rms
over an area of radius 3 pixels; the rms is determined from
sampling across the image.

\section{Stars} \label{sec-stars}

\subsection{Integrated properties}

The [3.6] image offers an alternate perspective on the stellar
populations of a galaxy compared to the optical. In principle
the near-IR is dominated by old stars and low mass stars of all
ages, although young red supergiants also contribute
from pockets of recent star formation. 
The integrated galactic [3.6] magnitudes are reported in Table \ref{tab-int}
as an absolute magnitude $M_{[3.6]}$ using the distances in Table \ref{tab-sample}.
Several integrated colors and $M_V$ are plotted in Figure \ref{fig-mch1}
as a function of
$M_{[3.6]}$. There is a strong correlation between the two absolute
magnitudes $M_V$ and $M_{[3.6]}$, as one would expect.
In terms of the colors there are correlations between $M_{[3.6]}$
and (\bv)$_0$ and between $M_{[3.6]}$ and
$V-$[3.6], and these correlations are in the sense
that more luminous galaxies are redder. The change
in (\bv)$_0$ is not large, however; over 8 mag in $M_{[3.6]}$
(\bv)$_0$ reddens by only $\sim$0.2 mag and the scatter is comparable
to the change. The change in $V-$[3.6] is larger; $V-$[3.6] changes by 
1.25 mag over that same interval.

The color [3.6]$-$[4.5]$_*$, on the other hand, is constant among
most of the galaxies at an average value of
$-0.08\pm0.18$. 
This color is comparable
to that of an early K star (Pahre \et\ 2004, Willner \et\ 2004,
Reach \et\ 2005).
In a study of a sample of galaxies spanning the Hubble sequence,
Pahre \et\ (2004) found that [3.6]$-$[4.5] becomes redder with
later Hubble type. The change is small, only $\sim$0.15 mag
from early to late-type galaxies, but it reaches about $\sim$0
at T$=10$ (Im galaxies). The small difference between their
[3.6]$-$[4.5] color and ours for Im galaxies could be due to
the fact that their [4.5] magnitude includes the hot dust component.
Therefore, their [3.6]$-$[4.5] color should be a little bit
redder than ours, as is observed.
Furthermore, the change in [3.6]$-$[4.5] color that they observe
could be due, at least in part,
to the increasing contribution of hot dust in
later type galaxies.
The average color of our sample is similar to that observed for the disk 
of M81 by Willner \et\ (2004; see their Figure 2).

The fact that the integrated stellar color
[3.6]$-$[4.5]$_*$ is constant from
galaxy to galaxy suggests simply that the old and low mass 
stellar component
of galaxies have nearly the same colors.
By contrast, that $V-$[3.6] changes more steeply with galactic absolute magnitude
is an indication of the relative importance of the young stellar component
to the older, low mass component.
Irregular galaxies are blue in optical colors primarily
because of the dominance of the young stellar population (and partly
due to low metallicity). Therefore, it makes sense that $V-$[3.6]
should reflect this. 

\subsection{Large-scale Morphology}

In some spirals, stellar
structures, such as bars, become more recognizable in the
infrared, perhaps
because they are formed from old stars or dust obscures their
features in the optical. 
In our galaxy sample, however, the large-scale morphology that we see
in the [3.6] or [4.5]$_*$ stellar images is the same as that which
we see in the optical. 
There are no hidden bars, and bars
that are present look the same in the infrared as in the optical
(DDO 154, NGC 3738, NGC 4214, NGC 4449, NGC 6822).
The stellar populations that are emphasized
may be different in different passbands, but the structure
of the galaxy is the same
for different stellar populations in Im galaxies.
One exception is IC 10 that lies at low Galactic latitude;
a Galactic dust lane obscures
the western part of the galaxy in the optical and the stars
both in IC 10 and in the Milky Way behind the dust lane
become apparent in the 3.6 $\mu$m image (see
Figure \ref{fig-regions}).
Another exception is NGC 3738. This Im galaxy with intense
central star formation looks similar in $V$ and [3.6],
but is less round in $J$ at outer isophotes (Hunter \& Elmegreen 2006).

\subsection{Embedded sources}

On a small scale there are some differences between
the [3.6] and $V$ band that are apparent in 4 of the galaxies:
IC 10, NGC 3738, NGC 4214, and NGC 4449.
These four galaxies are distinguished from the rest of the sample
by intense star formation; and in three of these systems 
star formation is nearly galaxy-wide (in NGC 3738 the star
formation is concentrated to the center of the galaxy).
These galaxies are shown in Figure \ref{fig-regions} where the
3.6 \micron\ and $V$ images have been combined in false color
and regions of 3.6 $\mu$m excess have been identified
in different passbands.

In IC 10 the main source of excess 3.6 \micron\ emission is
along the southeast edge of the highest surface brightness
part of the galaxy. In the optical there is an obvious dust
lane running northeast to southwest there, and we see enhanced diffuse
3.6 \micron\ emission from this region (numbered D1 in Figure 
\ref{fig-regions}). Along the northwest edge of this 
dust lane there is a region of star formation, prominent in \ha.
In the 3.6 \micron\ image there are 3 knots, all resolved with respect
to a point source. Two of these knots are visible in $V$ (numbered
1 and 3 in Figure \ref{fig-regions}); the third is not seen in $V$
(numbered 2 in Figure \ref{fig-regions}).
On the southeast edge of the dust lane, there is an unresolved
point source apparent at 3.6 \micron\ and not in $V$
(numbered 4 in Figure \ref{fig-regions}).

In NGC 3738, we see a source that is bright in 3.6 \micron\ in the
northern part of the central star-forming complex
(numbered 1 in Figure \ref{fig-regions}). This source
consists of two resolved knots. The western of the two
knots is associated with bright \ha\ emission and faint
$V$-band emission. The eastern of the two knots has
fainter \ha\ emission and is not visible in $V$.

In NGC 4214 the most interesting excess 3.6 \micron\ emission 
is associated with an arc of
star forming regions in the southern part of the galaxy.
Region 1 consists of two knots and region 2 is one knot,
all resolved with respect to a point source.
They are seen in both 3.6 $\mu$m and $V$, but there appears
to be excess 3.6 $\mu$m emission in and around these knots. 
Region D1 is a diffuse extension southward of knot 1
seen only in 3.6 $\mu$m. In addition to this star forming
region, diffuse excess 3.6 $\mu$m emission is associated
with a dust patch in the northwest part of the galaxy
(numbered D2 in Figure \ref{fig-regions}). We note that
a similar dust patch obvious in the optical located
at 12$^{\rm h}$ 15$^{\rm m}$ 39\fs3, 36\arcdeg\
19\arcmin\ 22\arcsec\ (J2000.0) does not show excess
3.6 $\mu$m emission.

NGC 4449 contains numerous regions of excess 3.6 $\mu$m emission.
Some of this emission is filamentary and suggests a
nebular origin. The similarity between the 3.6 \micron\ image and the
non-star-subtracted 4.5 \micron\ image suggests that some of the emission
at 3.6 \micron\ could be from hot dust or the 3.3 $\mu$m PAH feature.
Several regions of particular note are circled in Figure
\ref{fig-regions}. Region 1, located near the galaxy center,
is a resolved knot in 3.6 $\mu$m and a dust patch in the optical.
Regions D1 and D2 are diffuse 3.6 $\mu$m emission regions associated
with optical dust patches.
Region D3, also diffuse, 
lies along the outer edge of a large star forming complex 
in the northern part of the galaxy.

We have measured the brightnesses of the regions identified
in Figure \ref{fig-regions} in $UBV$, 3.6 $\mu$m, and
4.5 $\mu$m, where possible, and the photometry is given
in Table \ref{tab-regions}. Numbered regions are resolved knots
in 3.6 $\mu$m, and are photometered using circular apertures with
background subtraction determined from 
an annulus just beyond the aperture. The radius of the
aperture is given in Table \ref{tab-regions}. 
Regions designated with a ``D'' in the table are diffuse regions
that were photometered using a polygon whose shape was guided
by the $V/3.6$ \micron\ image. For the diffuse regions, there was no
background subtraction. The $UBV$ photometry was calibrated using
the data of Hunter \& Elmegreen (2006); the IRAC photometry
was calibrated as discussed in \S\ref{sec-data}.
In Table \ref{tab-regions} we also give the 3.6 $\mu$m surface brightness of
each region in the aperture. Since the calibration of the [3.6] image is 
relative to Vega, we 
report the surface brightness in terms of the luminosity
of Vega at 3.6 $\mu$m per square parsec, using
a distance to Vega of 7.6 pc. When the output of star clusters or
composite stellar populations
in the 3.6 $\mu$m
passband are better known, one can relate these surface brightnesses to 
a more appropriate stellar population.

From Table \ref{tab-regions} one can see that most of 
the [3.6]$-$[4.5]$_*$ colors
are those of early K stars, similar to the integrated colors
of Im galaxies. The resolved knots have surface brightnesses
comparable to 30--500 stars like Vega per square parsec, while the diffuse regions
have surface brightnesses of 50--120 Vegas per square parsec.
Photometry of star forming complexes in 4 galaxies suggest that
a $V-$[3.6] color of 2.3$\pm$0.5 is typical of young regions.
Given that, the knots in IC 10 have total A$_V$ of roughly 4.4--6.8 mag,
and reddening-corrected $M_V$ of order $-9.5$ to $-9.9$. Thus, they are
potentially populous star clusters (Billett \et\ 2002), although
the reddening corrected $UBV$ colors do not make very much sense
compared to cluster evolutionary tracks.
The two knots in NGC 3738 have an A$_V$ of order 2.7 if the intrinsic 
$V-$[3.6] color is 2.3 mag. That would yield an $M_V$ of $-12.5$,
potentially placing it in the realm of a super star cluster.
Knots 1 and 2 in NGC 4214 would each have A$_V$ of 0.5 mag,
and $M_V$ of $-11.5$ and $-11.9$. Thus, they could also
be massive star clusters. The bright diffuse regions in NGC 4449 are probably
patches of stellar disk.

The presence of embedded luminous clusters in four of our galaxies
is consistent with the star formation seen in other tracers. The
fact that there are only a few of these clusters in small galaxies
with overall star formation rates of $\sim0.05$ M$_\odot$
yr$^{-1}$ kpc$^{-1}$ is not surprising. In IC 10, the SFR is 0.049
M$_\odot$ yr$^{-1}$ kpc$^{-1}$ normalized to the exponential disk
radius of $R_D^V=1.27^\prime$ or 0.37 kpc. This gives a total star
formation rate of 0.02 M$_\odot$ yr$^{-1}$. If, in an extreme
case, all of this star formation were in the form of populous
clusters, each having a mass of $10^4$ M$_\odot$, then to get one
such cluster still in the embedded phase requires an obscuring
time of $10^4/0.02=5\times10^5$ yr.  In fact, $10^4$ M$_\odot$
clusters are relatively rare. For the usual $M^{-2}dM$ cluster
mass function, the fraction of the total star formation rate from
a decade interval of cluster mass is
$\log(10)/\log(M_U/M_L)=1/\log(M_U/M_L)$ for upper and lower
cluster mass limits $M_U$ and $M_L$. Typically,
$\log(M_U/M_L)\sim4$ to 5, depending on how a cluster is defined.
Thus $\sim20$\% of the star formation rate can be in ``populous''
clusters like the embedded one in IC 10. This means the
obscuration time is $10^4/(0.02*0.2)=2.5\times10^6$ yr.
This is actually a reasonable timescale considering observations
of local obscured clusters (e.g. Lada \& Lada 2003).  The total
star formation rates in NGC 3738 and NGC 4214 are 0.09 M$_\odot$
yr$^{-1}$ and 0.13 M$_\odot$ yr$^{-1}$, significantly larger than
in IC 10, so it is not unreasonable that the largest embedded
clusters are also more massive than in IC 10, if the obscuration
time is about the same.  This follows from statistically sampling
a universal cluster mass function.  All of the other Im
galaxies without bright embedded clusters could have their most
massive clusters at later phases in their cloud break out, or they
could have lower total star formation rates.

\subsection{Surface photometry}

The optical radial light profiles of Im galaxies are 
those of exponential disks (Hodge 1971),
and we see this same form in the mid-IR.
The [3.6] or [4.5]$_*$ azimuthally averaged stellar surface photometry 
is shown for each galaxy
in Figure \ref{fig-pix}. The central surface brightnesses
$\mu_0$ and disk scale lengths $R_D$ derived from exponential disk fits
to the surface photometry are given in Tables \ref{tab-int} and
\ref{tab-int2}, and are plotted against the $V$-band measurements
in Figures \ref{fig-mu0} and \ref{fig-rd}.
The central surface brightnesses and disk scale lengths measured
from the IRAC images are clearly correlated with those measured
from $V$-band images. That the disk scale lengths are
similar measured at different wavelengths implies
that the structure of the disk is usually the same for old and low mass stars
as for younger and more massive stars. 

We have previously found complex $V$-band surface brightness
profiles in approximately one-quarter of our survey 
of Im and Sm galaxies (Hunter \& Elmegreen 2006). 
These profiles are double exponentials:
one exponential fits the surface brightnesses in the central regions 
of the galaxy disk and an exponential 
with a different slope fits the outer parts.
The most common type of profile is one in which the outer part
of the galaxy exhibits an exponential form that is steeper than
that in the central part of the galaxy. In a few systems the
inner profile was actually flat, and in one system the outer profile
was shallower than the one at smaller radii. 
In our IRAC sub-sample, we have 
4 galaxies that exhibit one
of these complex profiles: DDO 165, IC 10, NGC 3738, and VIIZw 403.
In Figure \ref{fig-pix} one can see that the profiles in
[3.6] are very similar to those in $V$ band: The breaks in the profiles
are obvious and occur at, or nearly at, the same radii, and the
slopes in the profiles before and after the breaks are similar.
Therefore, the breaks and changes in the radial surface density of
stars is the same in different stellar populations within the galaxy.

Azimuthally-averaged colors are also shown as a function of radius
for each galaxy in Figure \ref{fig-pix}.
We find that 68\% (13 of 19) of the Im galaxies have $V-$[3.6]
or $V-$[4.5]$_*$ colors that are constant with radius. Thus, the 
mixture of young and old stars is the same throughout the galaxy disks.
Both BCDs with
these colors show a gradient, at least in the galaxy center,
getting redder with distance from
the center of the galaxy. This is consistent with the central concentration
of intense star formation in these systems.
For [3.6]$-$[4.5]$_*$, the color is constant with radius in
85\% (11 of 13) of the Im galaxies and
all 3 of the BCDs. 

\section{Hot Dust} \label{sec-dust}

\subsection{Spatial distribution}

The star-subtracted [4.5]$_{\rm dust}$ image shows the distribution and
brightness of hot dust emission in a galaxy.
The distribution of the hot dust in our sample of galaxies 
is shown as contours superposed on \ha\ images in Figure
\ref{fig-pix}. We note that the hot dust is always
much less extensive than the PAHs as seen in 8.0 $\mu$m-band emission.
In fact, most, and probably all, regions of hot dust
are associated with bright \HII\ regions.
There are a few regions in several galaxies that are not
associated with \HII\ regions, but these are most likely foreground
or background objects and not associated with the object
galaxy at all.
\HII\ regions are where the concentrations of UV photons and dust
are located, so this implies that young, massive
stars are the source of heating of this component of the dust.

We have measured the \ha\ surface brightness limit at which
hot dust is detected 
$\Sigma_{H\alpha}^{\rm [4.5]_{dust}}$.
The average is $8\times10^{29}$ erg s$^{-1}$ pc$^{-2}$ for
the Im galaxies with a factor of 8 variation among galaxies.
The average for 2 BCDs is $11\times10^{29}$ erg s$^{-1}$ pc$^{-2}$,
with a factor of 1.4 variation.
The \ha\ surface brightness limits are plotted as a function
of oxygen abundance and area-normalized star formation rate
in the bottom panels of Figure \ref{fig-limits}.
There is no correlation with oxygen abundance, but there is
a clear correlation with star formation rate: as the star formation
rate goes down, the \ha\ surface brightness limit at which hot
dust is detected also goes down.
Thus, there is no single surface brightness that predicts where
hot dust will be found for all Im galaxies.

\subsection{Integrated properties}

We have detected
[4.5]$_{\rm dust}$
emission in 69\% (11 of 16) of the galaxies with observations
in both 3.6 $\mu$m and 4.5 $\mu$m passbands.
The integrated brightness of the hot dust emission in the 4.5 \micron\ passband
is given as an absolute magnitude in Table \ref{tab-int}.
The oxygen abundance, area-normalized
star formation rate, and integrated (\ub)$_0$ color are
plotted as a function of this absolute magnitude $M_{\rm [4.5]_{dust}}$
in Figure \ref{fig-dust}. 
There is a slight increase in the amount of hot dust
with increasing star formation rate (correlation coefficient
0.6). This is consistent with the observation discussed in the
previous section that hot dust in Im galaxies is primarily detected
in \HII\ regions above
a particular surface brightness level.
There is no correlation of $M_{\rm [4.5]_{dust}}$ 
with (\ub)$_0$. There is a slight correlation with oxygen abundance
(correlation coefficient 0.5), probably reflecting the 
luminosity-metallicity relationship seen for galaxies since there
is a correlation between $M_{\rm [4.5]_{dust}}$ and the stellar absolute
magnitude $M_{[3.6]}$.

\section{PAHs} \label{sec-pahs}

\subsection{Integrated properties}

The star-subtracted [8.0]$_{\rm PAH}$ passband image contains emission from
PAHs, as well as 
from warm dust (Rosenberg \et\ 2006). 
We have detected this emission in 73\% (16 of 22) of the galaxies
in our sample.
The integrated brightness in the non-stellar [8.0]$_{\rm PAH}$ image converted
to an absolute magnitude $M_{[8.0]_{\rm PAH}}$ is given in Tables \ref{tab-int}
and \ref{tab-int2} and plotted
against integrated $M_V$, several colors, the area-normalized star
formation rate, and oxygen abundance in Figure \ref{fig-mch4}.
PAH emission increases with increasing $V$-band (or 3.6 \micron) luminosity.
There is also a correlation between $M_{[8.0]_{\rm PAH}}$ and [5.8]$_{\rm PAH}$$-$[8.0]$_{\rm PAH}$ color
in the sense that, as the PAH emission in the [8.0]$_{\rm PAH}$ image increases,
it does so faster than the emission in [5.8]$_{\rm PAH}$ increases. 
No [5.8]$_{\rm PAH}$ emission is detected in galaxies with $M_{[8.0]_{\rm PAH}}$ below
$-17$. This represents a detection limit for 5.8 \micron\ non-stellar emission
in our galaxy sample.

In Figure \ref{fig-ch1mch4} we plot the color [3.6]$-$[8.0]$_{\rm PAH}$ against
the 8.0 \micron\ PAH brightness, area-normalized star formation,
and oxygen abundance. The color [3.6]$-$[8.0]$_{\rm PAH}$ is
a measure of the [8.0]$_{\rm PAH}$ PAH emission relative to the mid-IR
stellar emission.
Most of the galaxies in our sample have [3.6]$-$[8.0]$_{\rm PAH}$
colors in the range 0--3 mag. 
Hogg \et\ (2005)
found that star-forming galaxies have larger PAH emission relative
to stars, but that low luminosity systems have unusually low PAH-to-star
ratios.
In a study of galaxies spanning the Hubble sequence,
Pahre \et\ (2004) found that later type galaxies have higher PAH emission relative
to starlight. They measure [3.6]$-$[8.0]$_{\rm PAH}$$\sim$2 for types greater than
T$=$3.  In the disk of M81, Willner \et\ (2004) found a relatively 
constant [3.6]$-$[8.0]$_{\rm PAH}$
color with radius at a value of about 0.5.
The high star formation rate dwarf sample of Rosenberg \et\ (2006) has
colors 1--3 mag.
Thus, our sample covers the range of colors seen in other samples of
Im or low luminosity galaxies.

There is a correlation between PAH emission and star formation rate,
seen in both Figure \ref{fig-mch4} and Figure \ref{fig-ch1mch4}:
PAH emission increases relative to stellar emission
as the star formation rate per unit area
increases.
Roussel \et\ (2001) 
and F\"orster Schreiber \et\ (2004) found a correlation between
star formation rate and the 5--8.5 $\mu$m emission in spiral and starburst systems,
and Dale \et\ (2005) have 
suggested that the [8.0]$_{\rm PAH}$ PAH emission generally traces
the star formation rate with factors of 10--20 variations in galaxies.
Rosenberg \et\ (2006) also saw this correlation in their sample of 
starburst dwarfs.

The correlation of PAH emission with star formation rate
suggests that the stars that ionize the gas also
heat the PAHs. 
This is consistent 
with the spatial distribution
discussed in \S5.3 in which, for most galaxies, the PAH
emission is found in the brightest \HII\ regions. 
Although cool stars
can heat PAHs too (Uchida \et\ 1998, Li \& Draine 2002), 
in these galaxies the primary
source of heating are the concentrations of young, massive stars in 
\HII\ regions. 
The exceptions in our sample are
the galaxies with very intense, galaxy-wide star formation. In
those systems the PAH emission is much more extended.
But, the wide-spread nature of the star formation in
these systems, coupled with the porosity that comes from
concentrations of massive stars blowing holes in the ISM, allows UV photons
from the star-forming regions to escape to large distances. Thus,
the heating source for the PAHs is likely the same in all of the
galaxies in our sample.
However, the strong UV fluxes can also
destroy or ionize PAHs (Boulange \et\ 1988, Vestraete \et\ 1996, 
Tacconi-Garmon \et\ 2005).
Therefore, we would expect that very high spatial resolution imaging
would reveal that the PAH emission is actually located in the outer
parts of the \HII\ regions, as is customarily found 
(for example, Churchwell \et\ 2004, Marston \et\ 2004, Rho \et\ 2006).

However, Spoon (2003) has suggested that on galactic scales the PAH emission
better traces the B stars rather than sites of younger star formation.
In a study of mid-IR emission in M51, Calzetti \et\ (2005) state
that the UV photons that heat the PAHs come from recent past star
formation ($\leq$100 Myr old), as well as current star
formation ($\leq$10 Myr). Therefore,
the [8.0]$_{\rm PAH}$ PAH emission is correlated with star formation but not
directly proportional to the number of ionizing photons.
For our sample, in Figure \ref{fig-mch4} we see that the [8.0]$_{\rm PAH}$ PAH brightness
is independent of integrated galactic (\ub)$_0$. 
(\ub)$_0$ is dominated by the young stellar component of the galaxy, but
since it
contains a contribution from relatively older, redder B stars as well as OB
stars in star-forming regions, we might expect (\ub)$_0$ to become
bluer (up to the (\ub)$_0$ color of an O star)
as PAH emission increases if the only contribution to the PAH
heating were from OB stars in \HII\ regions. That we do not see
such a trend suggests that B stars outside of \HII\ regions are
also contributing to PAH heating in Im galaxies.
On the other 
hand, in Figure \ref{fig-ch1mch4} galaxies with a greater proportion of
PAH emission relative to stellar 3.6 \micron\ emission are in fact slightly
bluer in (\ub)$_0$. (There is no separate correlation between M$_{[3.6]}$
or M$_{[8.0]_{\rm PAH}}$ and (\ub)$_0$; see Figures \ref{fig-mch1} and
\ref{fig-mch4}). However, the correlation is shallow. 
All of this together suggests that 
young, massive stars in star-forming regions are a major source
of heating of PAHs in Im galaxies, but that B stars outside of star-forming
regions also contribute.

Engelbracht \et\ (2005) has found that the [8.0]$_{\rm PAH}$ PAH emission decreases
relative to the 24 $\mu$m dust emission for galaxies with metallicity
less than 1/3--1/5 solar (12$+\log$O/H$\leq 8.2$).
Thus, we expect most, if not all, of the galaxies in our sample to have
depressed PAH emission, and as a group this may be true. 
However, we have explored the relationship between metallicity and
PAH emission within this low metallicity sample.
In Figures \ref{fig-mch4} and \ref{fig-ch1mch4} we see that
galaxies with higher oxygen abundances have higher
[8.0]$_{\rm PAH}$ PAH emission and more PAH emission relative to that
of stars at 3.6 $\mu$m. 
Rosenberg \et\ (2006) also see a correlation between $M_{[8.0]_{\rm PAH}}$
and oxygen abundance for their star-forming dwarfs, 
and the slopes of our relationships are
very similar: We measure $0.08\pm0.02$, and they measure 0.07.
Rosenberg \et\ and
Hogg \et\ (2005) 
also see 
a slight trend with a lot of scatter in a plot of [3.6]$-$[8.0]$_{\rm PAH}$ with
oxygen abundance for low luminosity systems.

\subsection{Surface photometry}

In Figure \ref{fig-pix} we explore the variation of non-stellar
[5.8]$_{\rm PAH}$$-$[8.0]$_{\rm PAH}$ color with radius through 
azimuthally averaged surface photometry 
for each galaxy. We see that in 67\% (4 of 6) of the Im galaxies
and one of two BCDs
this color is constant with radius.
In a study of the disk of M81, Willner \et\ (2004) found that the 
[5.8]$_{\rm PAH}$$-$[8.0]$_{\rm PAH}$ color of non-stellar emission varies between
about 0 and 0.5 with radius.

\subsection{Two-dimensional spatial distribution}

We illustrate the distribution of PAH emission relative to
the star formation activity as seen in nebular \ha\ emission
in Figure \ref{fig-pix}. 
There we have superposed a low surface brightness contour of [8.0]$_{\rm PAH}$ PAH
emission 
on our \ha\ image.
We see that in most of the galaxies the PAH
emission is associated with the brighter \HII\ regions.
This is also what we found from a study of a smaller sample of
Im galaxies with {\it ISO} (Hunter \et\ 2001).
As discussed above, this implies that the young, massive stars
are the primary heating source for the PAH and dust
emission in the [8.0]$_{\rm PAH}$ passband in Im galaxies.
This is in contrast to the study of M81 by Willner \et\ (2004)
who found that [8.0]$_{\rm PAH}$ PAH emission corresponds best with dust
lanes in that spiral disk.

The average \ha\ surface brightness limit for detecting PAH emission, 
excluding galaxies with
extended emission, is $3\times10^{29}$ ergs s$^{-1}$ pc$^{-2}$
with a factor of 12 variation for the Im galaxies. For the
two BCDs, the average limit is $10\times10^{29}$ ergs s$^{-1}$ pc$^{-2}$
with a factor of 10 variation.
These limits are plotted as a function of oxygen abundance and
area-normalized star formation rate in the top panels of Figure \ref{fig-limits}.
Like the hot dust \ha\ surface brightness limits, the lower the star
formation rate, the lower the \ha\ surface brightness limit
at which PAH emission is detected.
However, unlike that for hot dust emission, there is a correlation
between the limits for PAH emission and oxygen abundance: 
Galaxies with higher metallicities have lower \ha\ surface brightness
limits for PAH emission.

Four galaxies in our sample have PAH emission that is much more extensive
than location in a few discrete \HII\ regions. These galaxies
are IC 10, NGC 3738, NGC 4214, and NGC 4449. Not surprisingly,
these galaxies are all characterized
by intense, galaxy-wide star formation as well.
In these galaxies the PAH emission is highly extended, even
into the outer parts of the galaxy. NGC 6822 also has 8.0 \micron\
emission outside of \HII\ regions, although its emission is more 
localized and not as extensive as those in the other four galaxies.

Tacconi-Garman \et\ (2005) found no correlation between
star-forming regions and PAH emission on local scales
in a study of two starburst systems. 
This is not the case in our sample where there is a fairly high
degree of correlation between \ha\ and PAH emission although the
relative strengths vary.
For four of the galaxies with extended PAH emission
we have plotted 
azimuthally averaged radial profiles of the \ha\ emission
relative to the [8.0]$_{\rm PAH}$ PAH emission
in Figure \ref{fig-hadivch4}.
(Because of Milky Way cirrus, we do not have measurements in 
the 8.0 \micron\ band for NGC 6822).
We see that in NGC 3738, NGC 4214,
and the inner part of NGC 4449
the ratio of \ha\ to PAH emission remains relatively
constant with radius.
In the outer part of NGC 4449 and over most of IC 10 this
ratio decreases with radius;
the average PAH emission is stronger than we would expect for the \ha\ emission
at those radii.

We also see this on a local level.
Figure \ref{fig-color} displays the \ha\ and [8.0]$_{\rm PAH}$ PAH images
in false color, with \ha\ in green and [8.0]$_{\rm PAH}$ in red, for the
5 galaxies with extra-\HII\ region PAH emission. One can see that 
most of the PAH emission is correlated with
\ha\ emission, but not always. There are places within the
galaxies where PAH emission is found without \ha\ 
or the \ha\
brightness is very low. 
The center of NGC 6822 is an example of high PAH emission with almost
no \ha\ emission. 
The NGC 6822 field is contaminated by Galactic cirrus, but
the region in question is higher in surface brightness and smaller
in size than the large-scale pattern one sees across the image.
Spoon (2003) and Calzetti \et\ (2005) have suggested that B stars no
longer associated with \HII\ regions can be a major source of heating
of the PAHs in galaxies. 
This could be the case for this region in NGC 6822, because that
region of the disk is slightly redder than its surroundings. The
lack of star formation means there are few dense molecular clouds.
There can still be PAH emission, however, if there are diffuse clouds
irradiated by local B stars and more remote OB stars. The
inner part of NGC 6822 could therefore be unusual in the sense
that most of the ISM is in a diffuse form that does not readily 
form stars.
The same could be true for IC 10 where the extended regions
of high PAH-to-\ha\ emission could be irradiated by local B
stars and distant OB stars. This is somewhat easier to understand
in IC 10 than in NGC 6822 because IC 10 has many gas holes
through which UV radiation can travel long distances
(Wilcots \& Miller 1998).

\subsection{PAH emission in shells and supershells}

Figure \ref{fig-fil} illustrates the fact that PAH emission
is also found along the rims of gas shells and filaments. 
One can see this in IC 10 where the two gas shells to the northwest
of the highest surface brightness part of the galaxy
(\ha\ shells numbered 5 and 7 by Wilcots \& Miller 1998; see their Figure 11)
have PAH emission mixed with \ha\ emission. 
In fact these shells are more complete in 
the [8.0]$_{\rm PAH}$ image than in \ha. This is especially true of shell number
5 where the \ha\ emission is concentrated in \HII\ regions embedded
in the \HI\ shell, while the [8.0]$_{\rm PAH}$ image shows fainter extra-\HII\
region emission along the shell.
These shells, seen in \HI\ as well, are about 200-290 pc in diameter 
(for a distance of 1 Mpc; Wilcots \& Miller 1998)
and represent the mechanical energy input of concentrations of massive stars
to their surroundings.
Cannon \et\ (2005) also see PAH emission associated with a gas
shell in the Im galaxy IC 2574.

An even more extreme example is found in NGC 4449.
There one can see PAH emission associated with a supergiant gas
filament in the northwest part of the outer galaxy. 
These are filaments
5 and 6 of Hunter \& Gallagher (1992; and best seen in Figures 1 and 4 of
Hunter \& Gallagher 1997). The \ha\ filaments
form a coherent structure 1.7 kpc long, and
filament 6 falls along the inner edge of a 1.8 kpc diameter \HI\ supershell.
PAH emission is found in broken patches along these filaments.
At the base of this supershell near the center of the galaxy ionized
gas is expanding at 30 \kms. These and other structures create a picture in
which intense star-forming events have created multiple supergiant
shells in the ISM of NGC 4449 (Hunter \& Gallagher 1997).
Thus, PAHs are not destroyed in the shell formation process, but rather
are swept up into shells and filaments along with the gas.
Tacconi-Garman \et\ (2005) suggest that this means that
galaxies are capable of enriching their haloes and the intergalactic
medium with dust.

The PAH emission is not always spatially coincident with the \HI\
or \ha\ emission in the shells. In IC 10's shell number 5 and the
northwest edge of shell number 7 (see Figure 12 of Wilcots \&
Miller 1998) the \HI, \ha, and PAH emission are spatially
coincident. However, along the northern edge of shell 7, the \ha\
emission is located along the inner edge of the \HI\ ridge while
the 8.0 \micron\ emission extends further into the ridge by 10--30
pc.  The 8.0 \micron\ appears brightest where the \ha\ is
brightest in this shell.  Also, in the far northeast, the \HI\
emission continues for a distance of at least twice the field of
view in Figure 11 with a column density of $\sim10^{21}$ cm$^{-2}$
(Wilcots \& Miller 1998), but there is little 8.0 \micron\
emission in the northeastern part of our field of view. On the
other hand, in the western part of Figure 11, there is a lot of
dispersed 8.0 \micron\ covering a broad area where the \HI\ is
lacking.

In NGC 4449 the \ha\ and PAH emission are spatially coincident
in filament 5,
to within our ability to measure the positions (about 2\arcsec$=$38 pc),
but both lie along the inner edge of the \HI\ ridge, offset
about 265 pc (14\arcsec) from the ridge center.
(The FWHM of the \HI\ beam is 15\arcsec).
Along the \HI\ supershell, the diffuse \ha\ and PAH
emission along the northwest edge appear to be spatially coincident,
although the center of such faint, diffuse emission is difficult
to accurately determine. The center of the HI ridge is offset further out
by 190 pc or 0.3$R_{\rm shell}$. A different situation pertains
along the southwest (bottom) edge of the \HI\ supershell:
The \HI\ and [8.0]$_{\rm PAH}$ emission are spatially coincident while
the \ha\ lies along the inner \HI\ edge, offset by 
110-210 pc (0.15-0.3R$_{\rm shell}$) from the center of the
ridge.

These differences between 8 \micron\ and \HI\ could be from a
combination of several effects. First, the 8.0 \micron\ emission
from PAHs requires both gas and starlight, so there can be a
decrease in 8.0 \micron\ relative to \HI\ where the radiation field
decreases in the presence of neutral gas. For example, in IC 10,
the decrease in 8.0 \micron\ compared to \HI\ in the northern part
of shell 7 may occur because of extinction inside the ridge.
Similarly, the relative decrease of 8.0 \micron\ in the northeast
of IC 10 may occur because of the great distance from the
radiation source, which is concentrated in the center of the
galaxy. The concentration of 8.0 \micron\ to the irradiated parts
of \HI\ shells in NGC 4449 is probably a similar effect. A second
process that could change the relative emission strengths of 8.0
\micron\ and \HI\ is diffuse ionization. The extended diffuse 8.0
\micron\ to the west of IC 10 (see above) corresponds to a giant \HI\
hole that could be mostly diffuse ionized gas; there is a bright Wolf-Rayet
star inside this \HI\ hole 
(Massey \et\ 1992; Wilcots \& Miller 1998).  The
PAHs would
then be dispersed and irradiated in the ionized gas. These two effects
should
cause a different relative abundance of ionized versus neutral PAH
emission, which might be detectable in IR spectra.

\section{Summary} \label{sec-summary}

We have examined mid-IR images of 22 irregular galaxies
obtained from the {\it Spitzer} archives. The galaxies are representative
of normal systems of the Im, BCD, and Sm classes and span the range
of luminosities and star formation rates found in larger samples.
From the [3.6] and [4.5] images we extract information about the structure
of the stellar disk and its stellar populations.  
The star-subtracted [4.5]$_{\rm dust}$ image is used to examine the distribution
and integrated emission from hot dust.
The nebular [5.8]$_{\rm PAH}$ and [8.0]$_{\rm PAH}$ images are used to explore the properties of
PAH and hot dust emission. The two-dimensional distributions,
integrated magnitudes and colors, and azimuthally averaged surface
photometry are compared to optical images and measurements, 
including \ha\ images that trace the current star formation activity
and location of luminous, hot stars.

From the stellar images, we find that generally there are no hidden bar structures
in IRAC bands. 
Four of the Im galaxies have optically-hidden bright IR sources, with $M_V$
at least as bright as 
$-9.5$ to $-12.5$ mag, placing them in the range for populous and
super star clusters. 
Abrupt changes in some radial stellar density profiles seen in the optical,
suggesting double-exponential disks, are also seen in the near-IR.
For most Im and all BCD and Sm galaxies,
the overall structure of the galaxy is the same in the different passbands.
Furthermore, the colors suggest that in most Im galaxies the
mixture of young and old stars is the same throughout the galaxy disk.

The hot dust seen in the star-subtracted [4.5]$_{\rm dust}$ image is detected
in 2/3 of the galaxies. The hot dust is found only in the 
higher surface brightness \HII\ regions, 
where the concentrations of UV photons as well as dust are located.
This implies that young, massive
stars are the source of heating of this component of the dust in 
Im and BCD galaxies.
Furthermore, galaxies with 
higher area-normalized star formation rates have more dust emission.

PAH plus dust emission in the star-subtracted [8.0]$_{\rm PAH}$ image
is detected in nearly 3/4 of our sample of galaxies.
As the area-normalized star formation rate increases, the PAH emission
increases relative to the starlight. Furthermore, except for a few
galaxies, the PAH emission is found only in the brightest \HII\ regions.
The exceptions are galaxies
with intense, wide-spread star formation, and in
those systems the PAH emission is much more extended, even to 
the outer parts of the galaxy.
The correlation with star formation activity
suggests that, on average, the concentrations of young, massive
stars that ionize the gas also primarily heat the PAHs in Im-type galaxies.
However, on a local level, the ratio of PAH emission to \ha\ emission
is not constant.

We see PAH emission associated with
the rims of gas shells and super-giant filaments that have resulted
from the intense mechanical energy input from concentrations
of massive stars.
This implies that PAHs are not destroyed in the shell formation process, 
but are swept up into the shells and filaments along with the gas
even to the outer parts of a galaxy.

\acknowledgments
E. M. would like to thank Kathy Eastwood and the 2005
Research Experience for Undergraduates program at Northern Arizona
University which is funded by the National Science Foundation (NSF)
under grant AST-0453611.
We appreciate educational exchanges about IRAC with M. Pahre, S. Willner, 
and M. Ashby. 
We thank M. Fitzpatrick for help with an IRAF problem.
Funding for this work was provided to D. A. H. by the Lowell Research
Fund and the NSF through grant AST 02-04922.
This publication makes use of data products from the Two Micron All Sky 
Survey, which is a joint project of the University of Massachusetts 
and the Infrared Processing and Analysis Center (IPAC)/California Institute 
of Technology, funded by the National Aeronautics and Space 
Administration (NASA) and NSF.
This research has also made
use of the NASA/IPAC Extragalactic Database (NED) which is
operated by the Jet Propulsion Laboratory, California Institute of
Technology, under contract with NASA.

Facilities: \facility{Spitzer Space Telescope}, \facility{Lowell Observatory}.

\clearpage

\begin{figure}
\caption{Images and surface photometry for all of the galaxies 
in this study, two rows per galaxy and two galaxies per page. 
The images all show the same field of view for each galaxy.
{\it Top left:} The geometrically transformed [3.6] or [4.5] image or
the logarithm of the image. The contour is of an outer $V$-band isophote
(Hunter \& Elmegreen 2006).
{\it Top middle:} An \protect\ha\ image (Hunter \& Elmegreen 2004), 
if the galaxy has \protect\ha\ 
emission, is shown with a contour of [8.0]$_{\rm PAH}$ PAH emission (with
stars removed as discussed in \S2).
If the galaxy has no detectable PAH emission, the contour is the
same $V$-band contour shown in the upper left image.
{\it Bottom left:} An \protect\ha\ image, if the galaxy has \protect\ha\ 
emission, is shown with a contour of [4.5]$_{\rm dust}$ 
hot dust emission (stars
removed as discussed in \S2), if the galaxy has detectable hot dust emission.
{\it Bottom middle:} 
The surface photometry
in $V$, \ha, [3.6] or [4.5]$_*$, and [8.0]$_{\rm PAH}$ PAH emission,
if detected, as a function of radius is plotted here. 
The scale on the left vertical axis is for $V$ band, and the scale on the
right vertical axis is for \protect\ha. The zero points 
of the IRAC surface photometry have been adjusted so that plots of
the different quantities are easily distinguished, but
the scales 
have all been set
so that they cover the same logarithmic interval.
The solid lines are
fits to the exponential disks in $V$, $J$ if observed,
and [3.6] or [4.5]$_*$.
{\it Right:} Plot of the colors as a function
of radius from the azimuthally-averaged annular photometry.
[The first of 11 parts is shown here.]
\label{fig-pix}}
\end{figure}


\clearpage

\begin{figure}
\epsscale{0.70}
\caption{Integrated absolute $V$ magnitude and several colors
plotted as a function of absolute [3.6] magnitude.
The dashed lines are fits to the data:
{\it Top panel:} $M_V = (-0.83\pm0.63) + (0.83\pm0.04){M}_{[3.6]}$;
{\it Second panel:} (\bv)$_0 = (-0.08\pm0.19) - (0.02\pm0.01){M}_{[3.6]}$;
{\it Third panel:} (\ub)$_0 = -0.31\pm0.16$;
{\it Fourth panel:} $V-$[3.6] $= (-0.42\pm0.47) - (0.15\pm0.03){M}_{[3.6]}$;
{\it Bottom panel:} $[3.6]-[4.5]_* = -0.08\pm0.18$ (highly discrepant point
is not included in average).
[4.5]$_*$ is the stellar contribution to the 4.5 $\mu$m image; separation of
the dust and stellar components is discussed in \S2.
\label{fig-mch1}}
\end{figure}

\clearpage

\begin{figure}
\epsscale{1.0}
\caption{Comparison of 3.6 $\mu$m images with other
passbands for the 4 galaxies
with regions of excess 3.6 $\mu$m emission: IC 10, 
NGC 3738, NGC 4214, and NGC 4449.
There are 5 panels for each galaxy. The 4 panels in black and white
show the same field of view; the color panel generally
shows a somewhat larger
field of view.
{\it Top right:} False color images comparing $V$ ({\it green}) and 3.6 $\mu$m
({\it red}). Regions of interest are circled.
{\it Top left:} $V$/3.6 $\mu$m image. Photometered 
regions are circled and numbered
as in Table \protect\ref{tab-regions}, and the positions of the numbers
and circles are retained in the other 3 black and white panels. 
White in this image represents excess 3.6 $\mu$m emission.
Some of the white regions in IC 10 and NGC 4214 are stars that were edited
from the $V$-band image and not the 3.6 $\mu$m image.
{\it Top middle:} \protect\ha\ image from Hunter \& Elmegreen (2004).
The stellar continuum has been removed.
{\it Bottom left:} 3.6 $\mu$m image. Foreground stars and background 
objects have been
removed from the NGC 3738 3.6 $\mu$m image.
{\it Bottom middle:} $V$-band image. Foreground stars and background galaxies
have been removed from the images except for IC 10.
\label{fig-regions}}
\end{figure}



\clearpage

\begin{figure}
\epsscale{0.80}
\caption{Central surface brightnesses $\mu_0$ from fits to
the IRAC surface photometry compared to that measured
in $V$ band (Elmegreen \& Hunter 2006). Galaxies with [3.6]
images are shown in the upper panel, and those with only
[4.5] images are shown in the bottom panel.
The dashed lines are fits to the points:
{\it Top panel:} $\mu_0^{[3.6]} = (-3.60\pm1.84) + (1.07\pm0.08)\mu_0^{V}$;
{\it Bottom panel:} $\mu_0^{[4.5]_*} = (-0.86\pm2.46) + (0.95\pm0.11)\mu_0^{V}$.
[4.5]$_*$ is the stellar contribution to the 4.5 $\mu$m image; separation of
the dust and stellar components is discussed in \S2.
The BCD (HS 0822$+$3541) in the top panel has been left
out of the fit.
\label{fig-mu0}}
\end{figure}

\clearpage

\begin{figure}
\epsscale{1.0}
\caption{Disk scale length measured from the IRAC [3.6] or [4.5]$_*$ image
$R_D^{[3.6]}$ or $R_D^{[4.5]_*}$ plotted against the disk
scale length measured from the $V$-band surface photometry
$R_D^V$ (from Hunter \& Elmegreen 2006). 
[4.5]$_*$ is the stellar contribution to the 4.5 $\mu$m image; separation of
the dust and stellar components is discussed in \S2.
For galaxies with double exponential profiles, one scale length
is plotted as a solid symbol and the second one is plotted as an
open symbol.
The dashed line delineates a one-to-one correspondence and
is not a fit to the data.
\label{fig-rd}}
\end{figure}

\clearpage

\begin{figure}
\epsscale{0.50}
\caption{\protect\ha\ surface brightness limits at which hot dust measured
in [4.5]$_{\rm dust}$ images
({\it Bottom panels})
and [8.0]$_{\rm PAH}$ PAH emission ({\it Top panels}) are detected in our sample galaxies are
plotted against oxygen abundance and area-normalized star formation rate.
[4.5]$_{\rm dust}$ is the hot dust contribution to the 4.5 $\mu$m image; 
separation of
the dust and stellar components is discussed in \S2.
The units of $\Sigma_{H\alpha}$ are ergs s$^{-1}$ pc$^{-2}$.
The oxygen abundance and star formation rates are integrated values
over each galaxy.
No limits were measured in galaxies 
with highly extended PAH emission. 
Dashed lines are fits to the data:
{\it Top left:} log $\Sigma_{H\alpha}^{[8.0]} = (41.78\pm2.52) - (1.55\pm0.32)
(12+\log {\rm O/H})$;
{\it Top right:} log $\Sigma_{H\alpha}^{[8.0]} = (30.20\pm0.50) + (0.28\pm0.24)
(\log \rm{SFR/area})$;
{\it Bottom left:} log $\Sigma_{H\alpha}^{\rm [4.5]_{\rm dust}} = 29.99\pm0.25$;
{\it Bottom right:} log $\Sigma_{H\alpha}^{\rm [4.5]_{\rm dust}} = (30.37\pm0.16) + (0.26\pm0.08)
(\log \rm{SFR/area})$.
\label{fig-limits}}
\end{figure}

\clearpage

\begin{figure}
\epsscale{0.80}
\caption{Integrated oxygen abundance, area-normalized star formation rate
(Table \protect\ref{tab-sample}),
and (\ub)$_0$ plotted as a function of the integrated brightness of
hot dust in the [4.5]$_{\rm dust}$ images.
[4.5]$_{\rm dust}$ is the hot dust contribution to the 4.5 $\mu$m image; 
separation of
the dust and stellar components is discussed in \S2.
The dashed lines are fits to the data:
{\it Bottom panel:} (\ub)$_0$ $= -0.44\pm0.21$;
{\it Middle panel:} $\log$ SFR/area $= (-5.50\pm1.71) - (0.26\pm0.11){M_{[4.5]_{\rm dust}}}$;
{\it Top panel:} $12 + \log {\rm O/H} = (6.48\pm0.89) - (0.10\pm0.06){M_{[4.5]_{\rm dust}}}$.
\label{fig-dust}}
\end{figure}

\clearpage

\begin{figure}
\epsscale{0.70}
\caption{Integrated $M_V$, PAH emission
color [5.8]$_{\rm PAH}$$-$[8.0]$_{\rm PAH}$, optical color
(\ub)$_0$, area-normalized star formation rate from Table 1, and oxygen 
abundance plotted as a function of the brightness of the PAH
emission in the [8.0]$_{\rm PAH}$ passband, reported as an absolute magnitude.
Open symbols with arrows are upper limits.
Dashed lines are fits to the data:
{\it Top panel:} $M_V = (-4.59\pm0.97) + (0.59\pm0.05){M}_{[8.0]_{\rm PAH}}$;
{\it Second panel:} [5.8]$_{\rm PAH}$$-$[8.0]$_{\rm PAH}$ $= (-1.96\pm0.33) - (0.16\pm0.02){M}_{[8.0]_{\rm PAH}}$
(fit to only the Im galaxies);
{\it Third panel:} (\ub)$_0 = -0.31\pm0.16$;
{\it Fourth panel:} $\log {\rm SFR/area} = (-6.96\pm1.70) - (0.27\pm0.09){M}_{[8.0]_{\rm PAH}}$
(fit only to $M_{[8.0]_{\rm PAH}} < -16$);
{\it Bottom panel:} $12+\log$O/H $= (6.48\pm0.37) - (0.08\pm0.02){M}_{[8.0]_{\rm PAH}}$.
\label{fig-mch4}}
\end{figure}

\clearpage

\begin{figure}
\epsscale{0.8}
\caption{Integrated brightness of [8.0]$_{\rm PAH}$ PAH emission, area-normalized
star formation rate from Table 1, and oxygen abundance plotted as a function of
the [3.6]$-$[8.0]$_{\rm PAH}$ color. This color represents the 
amount of stellar emission
relative to PAH emission.
Dashed lines are fits to the data:
{\it Top panel:} $M_{[8.0]_{\rm PAH}} = (-16.55\pm0.65) - (2.52\pm0.52)([3.6]-[8.0]_{\rm PAH})$;
{\it Second panel:} ${\rm (\protect\ub)_0} = (-0.33\pm0.05) + 
(-0.06\pm0.04)([3.6]-[8.0]_{\rm PAH})$ (excluding HS 0822$+$3541);
{\it Third panel:} $\log {\rm SFR/area} = (-2.31\pm0.30) + 
(0.66\pm0.22)([3.6]-[8.0]_{\rm PAH})$;
{\it Bottom panel:} $12+\log{\rm O/H} = (7.71\pm0.18) + (0.27\pm0.13)([3.6]-[8.0]_{\rm PAH})$.
\label{fig-ch1mch4}}
\end{figure}

\clearpage

\begin{figure}
\epsscale{1.0}
\caption{Azimuthally-averaged \protect\ha\ emission relative 
to [8.0]$_{\rm PAH}$ PAH emission for 4 of the galaxies with extended PAH emission.
The galaxy profiles have been adjusted vertically to
allow the different galaxies to be visually distinguished. 
NGC 6822 is not included because Milky Way cirrus prohibits quantitative
information from the 8.0 \micron\ image.
\label{fig-hadivch4}}
\end{figure}

\clearpage

\begin{figure}
\epsscale{0.85}
\caption{False color images of the galaxies with extra-\protect\HII\
region [8.0]$_{\rm PAH}$ PAH 
emission. \protect\ha\ is in green and the [8.0]$_{\rm PAH}$ PAH image is in
red. 
The field of view is 5.12\arcmin\ for IC 10, 2.08\arcmin\ for NGC 3738,
7.15\arcmin\ for NGC 4214, and 6.73\arcmin\ for NGC 4449.
The NGC 6822 image is 14.40\arcmin$\times$14.54\arcmin.
North is to the top and east to the left.
\label{fig-color}}
\end{figure}

\clearpage

\begin{figure}
\epsscale{0.8}
\caption{Deep displays of \protect\ha\ images of IC 10 and NGC 4449.
The contours are of [8.0]$_{\rm PAH}$ PAH emission. These images
illustrate the PAH emission associated with gas shells and supergiant
filaments. In NGC 4449 \protect\ha\ filaments 5 and 6 of Hunter \& Gallagher (1992, 1997),
also labelled here,
have PAH emission associated with it
in patches. There is also PAH emission associated with the supergiant \protect\HI\
shell of which filament 6 forms the ionized inner edge of a portion.
In IC 10 \protect\HI\ PAH emission delineates shells number 5 and 7 of Wilcots \& Miller (1998), labeled here near the centers of the holes.
\label{fig-fil}}
\end{figure}

\clearpage

\thispagestyle{empty}

                                                                                
%
                                                                                
                                                                                
\begin{deluxetable}{llcrccrccccc}                                               
\tabletypesize{\scriptsize}                                                     
\rotate                                                                         
\tablenum{1}                                                                    
\tablecolumns{12}                                                               
\tablewidth{0pt}                                                                
\tablecaption{Galaxy Sample \label{tab-sample}}                                 
\tablehead{                                                                     
\colhead{}                                                                      
& \colhead{}                                                                    
& \colhead{D\tablenotemark{b}} 
& \colhead{}                                                                    
& \colhead{}                                                                    
& \colhead{}                                                                    
& \colhead{}                                                                    
& \colhead{\protect\logsfrd\tablenotemark{f}}                                   
& \colhead{$\mu_0^{\rm V}\rm{(in)}$\tablenotemark{g}}                               
& \colhead{$R_D^{\rm V}\rm{(in)}$\tablenotemark{g}}                                 
& \colhead{$\mu_0^{\rm V}\rm{(out)}$\tablenotemark{g}}                              
& \colhead{$R_D^{\rm V}\rm{(out)}$\tablenotemark{g}} \\                             
\colhead{Galaxy}                                                                
& \colhead{Other Names\tablenotemark{a}}                                        
& \colhead{(Mpc)}                                                               
& \colhead{$M_V$\tablenotemark{c}}
& \colhead{($U-B$)$_0$\tablenotemark{d}}                                        
& \colhead{($B-V$)$_0$\tablenotemark{d}}                                        
& \colhead{O/H\tablenotemark{e}}                                                
& \colhead{(M\protect\solar/yr/kpc$^2$)}                                        
& \colhead{(mag arcsec$^{-2}$)}                                                 
& \colhead{(arcmin)}                                                            
& \colhead{(mag arcsec$^{-2}$)}                                                 
& \colhead{(arcmin)}                                                            
}                                                                               
\startdata                                                                      
\cutinhead{Im Galaxies}                                                         
CVnIdwA\dotfill  & UGCA 292                          &  4.1 & -12.65 & -0.54 &  0.21 & \nodata & -2.64 & 24.30 &  0.54 & \nodata & \nodata \\
DDO 50\dotfill   & PGC 23324,UGC 4305,HoII,VIIZw 223 &  3.4 & -16.61 & -0.41 &  0.22 &  7.83 & -1.83 & 22.19 &  1.11 & \nodata & \nodata \\
DDO 53\dotfill   & PGC 24050,UGC 4459,VIIZw 238      &  3.6 & -13.84 & -0.53 &  0.41 &  7.79 & -2.50 & 23.81 &  0.69 & \nodata & \nodata \\
DDO 63\dotfill   & PGC 27605,UGC 5139,HoI            &  3.8 & -14.73 & -0.18 &  0.20 & \nodata & -3.44 & 24.17 &  2.72 & \nodata & \nodata \\
DDO 69\dotfill   & PGC 28868,UGC 5364,Leo A,Leo III  &  0.8 & -11.67 & -0.29 &  0.29 &  7.27 & -3.00 & \nodata & \nodata & 23.01 &  0.81 \\
DDO 75\dotfill   & PGC 29653,UGCA 205,Sextans A      &  1.3 & -13.91 & -0.54 &  0.19 &  7.49 & -1.40 & \nodata & \nodata & 20.40 &  0.59 \\
DDO 154\dotfill  & PGC 43869,UGC 8024,NGC 4789A      &  4.3 & -14.51 & -0.32 &  0.30 &  7.70 & -2.60 & 23.06 &  0.55 & \nodata & \nodata \\
DDO 155\dotfill  & PGC 44491,UGC 8091,GR 8           &  2.2 & -12.53 & -0.45 &  0.28 &  7.68 & -1.50 & 21.72 &  0.23 & \nodata & \nodata \\
DDO 165\dotfill  & PGC 45372,UGC 8201,IIZw 499       &  4.8 & -15.69 & -0.21 &  0.23 & \nodata & -3.52 & 23.32 &  1.69 & 21.72 &  0.50 \\
DDO 187\dotfill  & PGC 50961,UGC 9128                &  2.5 & -12.95 & -0.18 &  0.30 &  7.62 & -2.64 & 22.05 &  0.28 & \nodata & \nodata \\
DDO 210\dotfill  & PGC 65367,Aquarius Dwarf          &  0.9 & -10.88 & -0.09 &  0.50 & \nodata & \nodata & 23.77 &  0.63 & \nodata & \nodata \\
DDO 216\dotfill  & PGC 71538,UGC 12613,Pegasus Dwarf &  0.9 & -13.29 & -0.03 &  0.66 &  7.93 & -4.15 & 23.44 &  1.67 & 21.26 &  1.03 \\
IC 10\dotfill    & PGC 1305,UGC 192                  &  1.0 & -17.11 & \nodata &  0.59 &  8.18 & -1.31 & 19.38 &  1.27 & 19.85 &  1.94 \\
M81dwA\dotfill   & PGC 23521                         &  3.6 & -11.73 & -0.33 &  0.27 & \nodata & \nodata & 23.32 &  0.25 & \nodata & \nodata \\
NGC 3738\dotfill & PGC 35856,UGC 6565,Arp 234        &  4.9 & -17.12 & -0.16 &  0.42 &  8.36 & -1.74 & 19.64 &  0.27 & 21.40 &  0.55 \\
NGC 4214\dotfill & PGC 39225,UGC 7278                &  2.9 & -17.56 & -0.38 &  0.37 &  8.43 & -1.10 & 20.42 &  0.86 & \nodata & \nodata \\
NGC 4449\dotfill & PGC 40973,UGC 7592                &  3.9 & -18.31 & -0.44 &  0.36 &  8.32 & -0.67 & 19.37 &  0.57 & 20.63 &  0.81 \\
NGC 6822\dotfill & PGC 63616,IC 4895,DDO 209         &  0.5 & -15.22 & -0.34 &  0.46 &  8.27 & -1.96 & 21.34 &  3.92 & \nodata & \nodata \\
\cutinhead{BCD Galaxies}                                                        
HS 0822$+$3542\dotfill &                             & 10.9 & -13.15 & -0.90 &  0.26 &  7.35 & -0.73 & 20.55 &  0.04 & \nodata & \nodata \\
NGC 1705\dotfill & PGC 16282                         &  5.0 & -16.25 & -0.52 &  0.28 &  8.21 & -0.82 & \nodata &  0.28 & \nodata & \nodata \\
VIIZw 403\dotfill & PGC 35286, UGC 6456              &  4.4 & -14.27 & -0.49 &  0.28 &  7.70 & -1.82 & 22.11 &  0.28 & 23.12 &  0.41 \\
  \\
  \\
  \\
  \\
  \\
  \\
  \\
\cutinhead{Sm Galaxies}                                                         
NGC 3109\dotfill & PGC 29128,UGCA 194,DDO 236        &  1.3 &  \nodata & \nodata & \nodata &  8.06 & -2.39 & \nodata &  3.27 & \nodata & \nodata \\
\enddata                                                                        
\tablenotetext{a}{Selected alternate identifications                            
obtained from NED.}                                                             
\tablenotetext{b}{Distances were taken from Hunter \& Elmegreen (2006),
and were determined from
variable stars or the tip of the Red Giant Branch, \\
if available in the literature.
Other distances
were determined from the radial velocity relative to the Galactic
standard of rest V$_{GSR}$ (RC3) \\
and a Hubble
constant of 65 \kms Mpc$^{-1}$. References for individual distances
are given by Hunter and Elmegreen.}
\tablenotetext{c}{$M_V$ are from Hunter \& Elmegreen (2006).}
\tablenotetext{d}{$UBV$ colors are from Hunter \& Elmegreen (2006)
except NGC 1705 whose colors are from de Vaucouleurs et al. (1991).}
\tablenotetext{e}{Oxygen abundance is 12$+$log(O/H). References are
given in Hunter \& Hoffman (1999) except                                        
DDO 69 and DDO 187 (Pilyugin 2001), \\                                          
DDO 154 (Kennicutt \& Skillman 2001),                                           
DDO 216 (Skillman et al.\ 1997),                                                
NGC 3109 (Richer \& McCall 1995),                                               
NGC 4449 (Talent 1980), \\                                                      
HS 0822$+$3542 (Kniazev et al.\ 2000), and                                      
NGC 1705 (Lee \& Skillman 2004).}                                               
\tablenotetext{f}{Integrated star formation rate normalized to the area         
of the galaxy within $R_D^V$, from Hunter \& Elmegreen (2004).}                 
\tablenotetext{g}{For surface brightness profiles with two parts,               
the first central surface brightness and disk scale-length are                  
the fit to the inner part of the profile, \\                                    
and the second set                                                              
of values are for the fit to the outer part of the profile,                     
from Hunter \& Elmegreen (2006).                                                
The NGC 3109 surface photometry \\                                              
was measured from an H$\alpha$ off-band image,                                  
NGC 1705 from the Digitized Sky Survey.
The radius at which the inner profile \\
makes the transition to the outer profile $R_{Br}$ are given
in Tables 2 and 3.} 
\end{deluxetable}                                                               

\clearpage

                                                                      
%
                                                                      
\begin{deluxetable}{lrrrrrrrccccc}
\tabletypesize{\scriptsize}                                           
\rotate                                                               
\tablenum{2}                                                          
\tablecolumns{13}
\tablewidth{0pt}                                                      
\tablecaption{Integrated Properties for Galaxies Observed in Four IRAC Passbands\tablenotemark{a}\label{tab-int}}  
\tablehead{                                                           
\colhead{}                                                            
& \colhead{}                                                          
& \colhead{}                                                          
& \colhead{}                                                          
& \colhead{}                                                          
& \colhead{}                                                          
& \colhead{}                                                          
& \colhead{}                                                          
& \colhead{$\mu_0^{[3.6]}\rm{(in)}$\tablenotemark{b}}                 
& \colhead{$R_D^{[3.6]}\rm{(in)}$\tablenotemark{b}}                   
& \colhead{$\mu_0^{[3.6]}\rm{(out)}$\tablenotemark{b}}                
& \colhead{$R_D^{[3.6]}\rm{(out)}$\tablenotemark{b}} 
& \colhead{$R_{Br}$\tablenotemark{b}} \\
\colhead{Galaxy}                                                      
& \colhead{$M_{[3.6]}$}                                               
& \colhead{[3.6]$-$[4.5]$_*$}
& \colhead{$V-$[3.6]}                                                 
& \colhead{$M_{[8.0]_{\rm PAH}}$}                                               
& \colhead{[5.8]$_{\rm PAH}$$-$[8.0]$_{\rm PAH}$}                                             
& \colhead{[3.6]$-$[8.0]$_{\rm PAH}$}                                             
& \colhead{$M_{\rm [4.5]_{dust}}$}                                          
& \colhead{(mag arcsec$^{-2}$)}                                      
& \colhead{(arcmin)}                                                  
& \colhead{(mag arcsec$^{-2}$)}                                      
& \colhead{(arcmin)}                                                  
& \colhead{(arcmin)}                                                  
}                                                                     
\startdata                                                            
\cutinhead{Im Galaxies}                                               
CVnIdwA\dotfill   &   -14.95 & -0.53 &  2.29 & $<$-13.84 & \nodata & \nodata & $<$-11.29 & 22.24 &  0.46 & \nodata & \nodata & \nodata \\
                       &   0.03 &  0.17 &  0.09 &   \nodata &  \nodata &  \nodata &   \nodata &  0.42 &  0.23 &  \nodata &  \nodata & \nodata \\
DDO 50\dotfill    &   -18.77 &  0.09 &  2.21 &    -18.99 &  0.97 &  0.22 &    -16.17 & 19.57 &  0.93 & \nodata & \nodata & \nodata \\
                       &   0.00 &  0.00 &  0.00 &   0.04 &  0.06 &  0.04 &   0.04 &  0.26 &  0.09 &  \nodata &  \nodata & \nodata \\
DDO 53\dotfill    &   -16.10 &  0.03 &  2.38 &    -16.90 &  0.72 &  0.81 &    -14.14 & 21.25 &  0.63 & \nodata & \nodata & \nodata \\
                       &   0.02 &  0.06 &  0.02 &   0.19 &  0.31 &  0.19 &   0.17 &  0.23 &  0.08 &  \nodata &  \nodata & \nodata \\
DDO 63\dotfill    &   -16.38 & -0.03 &  2.07 &    -15.09 & \nodata & -1.29 &    -12.20 & 22.10 &  2.75 & \nodata & \nodata & \nodata \\
                       &   0.01 &  0.02 &  0.02 &   0.89 &  \nodata &  0.89 &   0.95 &  0.13 &  1.38 &  \nodata &  \nodata & \nodata \\
DDO 154\dotfill   &   -15.82 & -0.07 &  1.66 & $<$-13.61 & \nodata & \nodata & $<$-10.92 & 21.37 &  0.54 & \nodata & \nodata & \nodata \\
                       &   0.01 &  0.03 &  0.02 &   \nodata &  \nodata &  \nodata &   \nodata &  0.12 &  0.05 &  \nodata &  \nodata & \nodata \\
DDO 165\dotfill   &   -17.47 &  0.12 &  1.92 &    -17.61 & \nodata &  0.14 & $<$-11.44 & 21.53 &  1.79 & 20.51 &  0.69 & 1.06 \\
                       &   0.00 &  0.01 &  0.01 &   0.15 &  \nodata &  0.15 &   \nodata &  0.04 &  0.19 &  0.10 &  0.03 & \nodata \\
DDO 187\dotfill   &   -14.35 &  0.27 &  1.68 & $<$-13.18 & \nodata & \nodata & $<$-10.71 & 20.23 &  0.26 & \nodata & \nodata & \nodata \\
                       &   0.01 &  0.03 &  0.01 &   \nodata &  \nodata &  \nodata &   \nodata &  0.45 &  0.06 &  \nodata &  \nodata & \nodata \\
IC 10\dotfill     &   -19.83 & -0.17 &  2.72 &    -21.87 &  1.54 &  2.03 &    -16.54 & 16.93 &  1.45 & 17.27 &  2.04 & 1.58 \\
                       &   0.00 &  0.00 &  0.00 &   0.00 &  0.00 &  0.00 &   0.00 &  0.08 &  0.16 &  0.07 &  0.11 & \nodata \\
M81dwA\dotfill    &   -13.36 & -0.37 &  1.63 &    -13.18 & \nodata & -0.18 & $<$-10.31 & 21.62 &  0.26 & \nodata & \nodata & \nodata \\
                       &   0.04 &  0.22 &  0.07 &   1.32 &  \nodata &  1.32 &   \nodata &  0.17 &  0.02 &  \nodata &  \nodata & \nodata \\
NGC 3738\dotfill  &   -19.62 & -0.04 &  2.50 &    -20.33 &  1.42 &  0.71 &    -15.12 & 17.50 &  0.29 & 19.91 &  1.04 & 0.88 \\
                       &   0.00 &  0.00 &  0.00 &   0.01 &  0.03 &  0.01 &   0.05 &  0.08 &  0.01 &  0.16 &  0.09 & \nodata \\
NGC 4214\dotfill  &   -20.01 & -0.07 &  2.54 &    -22.10 &  1.58 &  2.08 &    -16.52 & 17.23 &  0.60 & \nodata & \nodata & \nodata \\
                       &   0.00 &  0.00 &  0.00 &   0.00 &  0.01 &  0.00 &   0.04 &  0.09 &  0.02 &  \nodata &  \nodata & \nodata \\
NGC 4449\dotfill  &   -21.16 & -0.11 &  2.88 &    -23.74 &  1.76 &  2.57 &    -18.44 & 16.76 &  0.60 & \nodata & \nodata & 2.19 \\
                       &   0.00 &  0.00 &  0.00 &   0.00 &  0.01 &  0.00 &   0.01 &  0.08 &  0.02 &  \nodata &  \nodata & \nodata \\
NGC 6822\dotfill  &   -17.99 & -0.06 &  6.61 &     \nodata & \nodata & \nodata &    -12.59 & 18.96 &  4.46 & \nodata & \nodata & \nodata \\
                       &   0.00 &  0.00 &  0.00 &   \nodata &  \nodata &  \nodata &   0.06 &  0.07 &  0.56 &  \nodata &  \nodata & \nodata \\
\cutinhead{BCD Galaxies}                                              
HS 0822+3542\dotfill& -14.65 & -0.08 &  1.50 &    -15.16 & \nodata &  0.51 &    -13.09 & 19.86 &  0.05 & \nodata & \nodata & \nodata \\
                       &   0.09 &  0.25 &  0.09 &   0.76 &  \nodata &  0.76 &   0.77 &  0.14 &  0.00 &  \nodata &  \nodata & \nodata \\
NGC 1705\dotfill  &   -18.55 &  0.01 & \nodata &    -19.19 &  1.56 &  0.64 &    -15.43 & 17.84 &  0.25 & \nodata & \nodata & \nodata \\
                       &   0.00 &  0.00 &  \nodata &   0.02 &  0.05 &  0.02 &   0.03 &  0.15 &  0.01 &  \nodata &  \nodata & \nodata \\
VIIZw 403\dotfill &   -16.27 & -0.15 &  2.01 &    -16.97 &  1.09 &  0.69 &    -13.64 & 20.26 &  0.30 & 21.03 &  0.43 & 0.79 \\
                       &   0.01 &  0.04 &  0.02 &   0.12 &  0.24 &  0.12 &   0.20 &  0.05 &  0.01 &  0.05 &  0.01 & \nodata \\
\enddata                                                              
\tablenotetext{a}{Second line are the uncertainties for the quantities
in the first line. Values preceeded by ``$<$'' are upper limits.
Quantities [3.6] and [4.5]$_*$ \\
are stellar properties; 
[4.5]$_{\rm dust}$, [5.8]$_{\rm PAH}$, and [8.0]$_{\rm PAH}$ are measured from 
star-subtracted images.
Separation of the \\
dust and stellar components is discussed in \S2.}
\tablenotetext{b}{For surface brightness profiles with two parts,     
the first central surface brightness and disk scale length are        
the fit to the \\
inner part of the profile, and the second set 
of values are for the fit to the outer part of the profile.
$R_{Br}$ is the \\
radius at which this break occurs in the optical
from Hunter \& Elmegreen (2006).} 
\end{deluxetable}                                                     

\clearpage

                                                                                
%
                                                                                
\begin{deluxetable}{lrrrrccccc}
\tabletypesize{\scriptsize}
\rotate                  
\tablenum{3}             
\tablecolumns{10}        
\tablewidth{0pt}       
\tablecaption{Integrated Properties for Galaxies Observed in only [4.5] and [8.0]\tablenotemark{a}\label{tab-int2}}
\tablehead{                                                                     
\colhead{}                                                                      
& \colhead{}                                                                    
& \colhead{}                                                                    
& \colhead{}                                                                    
& \colhead{}                                                                    
& \colhead{$\mu_0^{[4.5]}\rm{(in)}$\tablenotemark{b}}                           
& \colhead{$R_D^{[4.5]}\rm{(in)}$\tablenotemark{b}}                             
& \colhead{$\mu_0^{[4.5]}\rm{(out)}$\tablenotemark{b}}                          
& \colhead{$R_D^{[4.5]}\rm{(out)}$\tablenotemark{b}} 
& \colhead{$R_{Br}$\tablenotemark{b}} \\
\colhead{Galaxy}                                                                
& \colhead{$M_{[4.5]_*}$}
& \colhead{$V-$[4.5]$_*$}
& \colhead{$M_{[8.0]_{\rm PAH}}$}
& \colhead{[4.5]$_*-$[8.0]$_{\rm PAH}$}
& \colhead{(mag arcsec$^{-2}$)}
& \colhead{(arcmin)}
& \colhead{(mag arcsec$^{-2}$)}
& \colhead{(arcmin)}
& \colhead{(arcmin)}
}                                                                               
\startdata                                                                      
\cutinhead{Im Galaxies}                                                         
DDO 69\dotfill    &   -13.14 &  1.76 & $<$-9.25 & \nodata & \nodata & \nodata & 21.39 &  0.99 & 1.16 \\
                       &   0.01 &  0.01 &   \nodata &  \nodata &  \nodata &  \nodata &  0.62 &  0.35 & \nodata \\
DDO 75\dotfill    &   -15.44 &  1.67 & $<$-10.92 & \nodata & \nodata & \nodata & 18.53 &  0.55 & 1.77 \\
                       &   0.01 &  0.02 &   \nodata &  \nodata &  \nodata &  \nodata &  0.24 &  0.03 & \nodata \\
DDO 155\dotfill   &   -14.17 &  1.83 &    -14.36 &  0.19 & 1\nodata &  0.24 & \nodata & \nodata & \nodata \\
                       &   0.01 &  0.01 &   0.21 &  0.21 &  0.51 &  0.06 &  \nodata &  \nodata & \nodata \\
DDO 210\dotfill   &   -12.60 &  1.91 & $<$-9.86 & \nodata & 21.94 &  0.70 & \nodata & \nodata & \nodata \\
                       &   0.06 &  0.06 &   \nodata &  \nodata &  0.15 &  0.08 &  \nodata &  \nodata & \nodata \\
DDO 216\dotfill   &   -15.38 &  2.34 &    -13.88 & -1.51 & 21.11 &  1.80 & \nodata & \nodata & 5.42 \\
                       &   0.01 &  0.01 &   0.30 &  0.30 &  0.01 &  0.02 &  \nodata &  \nodata & \nodata \\
\cutinhead{Sm Galaxies}                                                         
NGC 3109\dotfill  &   -17.68 & -0.52 &    -17.59 & -0.09 & 20.83 &  3.36 & \nodata & \nodata & \nodata \\
                       &   0.00 &  0.00 &   0.05 &  0.05 &  0.06 &  0.19 &  \nodata &  \nodata & \nodata \\
\enddata                                                                        
\tablenotetext{a}{Second line are the uncertainties for the quantities          
in the first line. Values preceeded by ``$<$'' are upper limits.
Quantities labeled [4.5]$_*$ are stellar properties;
quantities labeled [8.0]$_{\rm PAH}$ are measured from star-subtracted
images.
Separation of
the dust and stellar components is discussed in \S2. 
However, for the galaxies observed only in the [4.5] and [8.0] passbands,
the hot dust could not be separated from the stellar component
using the technique described there
and was assumed to be negligible.}
\tablenotetext{b}{For surface brightness profiles with two parts,               
the first central surface brightness and disk scale length are                  
the fit to the inner part of the profile, and the second set                    
of values are for the fit to the outer part of the profile.
$R_{Br}$ is the radius at which this break occurs in the optical
from Hunter \& Elmegreen (2006).}
\end{deluxetable}                                                               

\clearpage

                                                                      
%

\begin{deluxetable}{lcrrrrrrrccc}
\tabletypesize{\scriptsize}
\tablenum{4}
\tablecolumns{12}
\tablewidth{0pt}
\tablecaption{Photometry of 3.6 $\mu$m excess Regions\tablenotemark{a}\label{tab-regions}}  
\tablehead{
\colhead{}
& \colhead{}
& \colhead{R.A.}
& \colhead{Decl.}
& \colhead{$R$\tablenotemark{c}}
& \colhead{}
& \colhead{}
& \colhead{}
& \colhead{}
& \colhead{}
& \colhead{}
& \colhead{$\mu_{3.6}$} \\
\colhead{Galaxy}
& \colhead{Region\tablenotemark{b}}
& \colhead{(J2000.0)}
& \colhead{(J2000.0)}
& \colhead{(pc)}
& \colhead{$M_V$}
& \colhead{$B-V$}
& \colhead{$U-B$}
& \colhead{$M_{[3.6]}$}
& \colhead{$V-$[3.6]}
& \colhead{[3.6]$-$[4.5]$_*$\tablenotemark{d}}
& \colhead{($L^{Vega}_{3.6}$ pc$^{-2}$)}
}
\startdata
IC 10\dotfill    &  1 & 00 20 27.8 & 59 17 09 & 16.5 & -5.3 & 1.0 & \nodata & -12.2 & 6.9 & -0.07 & 150 \\
                 &    &            &          &      &  0.2 & 0.3 & \nodata &   0.0 & 0.2 &  0.01 &   1 \\
                 &  2 & 00 20 27.9 & 59 17 05 & 16.5 & -2.7 & 3.0 & \nodata & -11.8 & 9.1 & -0.08 & 110 \\
                 &    &            &          &      &  1.9 & 9.9 & \nodata &   0.0 & 1.9 &  0.01 &   1 \\
                 &  3 & 00 20 27.1 & 59 17 06 & 16.5 & -5.2 & 0.4 & \nodata & -12.5 & 7.3 & -0.07 & 200 \\
                 &    &            &          &      &  0.3 & 0.3 & \nodata &   0.0 & 0.3 &  0.01 &   1 \\
                 &  4 & 00 20 28.8 & 59 16 56 & 16.5 & \nodata & \nodata & \nodata & -11.9 & \nodata &  0.13 & 110 \\
                 &    &            &          &      & \nodata & \nodata & \nodata &   0.0 & \nodata &  0.01 &   1 \\
                 & D1 & 00 20 28.0 & 59 17 01 & 43   & \nodata & \nodata & \nodata & -13.6 & \nodata & -0.10 & 80 \\
                 &    &            &          &      & \nodata & \nodata & \nodata &   0.0 & \nodata &  0.01 &  1 \\
NGC 3738\dotfill &  1 & 11 35 48.6 & 54 31 32 & 107.8 & -9.8 & 1.4 & -2.9 & -14.5 & 5.0 & -0.04 & 30 \\
                 &    &            &          &       &  0.8 & 2.6 &  2.0 &   0.0 & 0.8 &  0.02  & 0  \\
NGC 4214\dotfill &  1 & 12 15 40.9 & 36 19 04 & 47.3  & -11.4 & 0.5 & -1.0 & -14.2 & 2.8 & -0.07 & 120 \\
                 &    &            &          &       &   0.0 & 0.0 &  0.0 &   0.0 & 0.0 &  0.01 &   0 \\
                 &  2 & 12 15 40.8 & 36 19 10 & 35.4  & -11.0 & 0.4 & -1.0 & -13.8 & 2.8 & -0.07 & 140 \\
                 &    &            &          &       &   0.0 & 0.0 &  0.0 &   0.0 & 0.0 &  0.01 &   1 \\
                 & D1 & 12 15 40.8 & 36 18 59 & 29.5  &  -7.7 & 0.4 & -1.0 & -12.2 & 4.5 & -0.19 &  46 \\
                 &    &            &          &       &   0.1 & 0.1 &  0.1 &   0.0 & 0.1 &  0.03 &   0 \\
                 & D2 & 12 15 36.8 & 36 20 02 & 84    & -11.6 & 0.4 & -0.1 & -14.4 & 2.8 & -0.03 & 46 \\
                 &    &            &          &       &   0.0 & 0.0 &  0.0 &   0.0 & 0.0 &  0.01 &  0 \\
NGC 4449\dotfill & 1 & 12 28 10.7 & 44 05 34 & 82    & \nodata & \nodata & \nodata & -16.9 & \nodata & -0.10 & 470 \\
                 &    &            &          &       & \nodata & \nodata & \nodata &   0.0 & \nodata &  0.01  &  1  \\
                 & D1 & 12 28 13.5 & 44 05 27 & 161   & -14.0 & 0.4 & -0.3 & -16.8 & 2.7 & -0.11 & 110 \\
                 &    &            &          &       &   0.0 & 0.0 &  0.0 &   0.0 & 0.0 &  0.01  &  0  \\
                 & D2 & 12 28 08.7 & 44 05 20 & 136   & -13.8 & 0.4 & -0.3 & -16.6 & 2.8 & -0.11 & 120 \\
                 &    &            &          &       &   0.0 & 0.0 &  0.0 &   0.0 & 0.0 &  0.01  &  0  \\
                 & D3 & 12 28 14.6 & 44 07 17 & 163   & -12.7 & 0.2 & -1.1 & -15.8 & 3.1 & -0.11 &  45 \\
                 &    &            &          &       &   0.0 & 0.0 &  0.0 &   0.0 & 0.0 &  0.02  &  0  \\
\enddata
\tablenotetext{a}{Second line are the uncertainties for the quantities
in the first line.} 
\tablenotetext{b}{Regions with ``D'' in their name are diffuse in the 3.6 $\mu$m image.}
\tablenotetext{c}{Radius of photometric aperture or, for polygons,
the radius of the circle with the same area as the polygon.}
\tablenotetext{d}{[4.5]$_*$ is the stellar contribution 
to the 4.5 $\mu$m image. Separation of
the dust and stellar components is discussed in \S2.}
\end{deluxetable}


\begin{thebibliography}

\bibitem[Allamandola et al.(1989)]{allamandola89} Allamandola, L. J.,
Tielens, G. G. M., \& Barker, J. R. 1989, ApJS, 71, 733
\bibitem[Billett et al.(2002)]{billett02} Billett, O. H., Hunter, D. A.,
\& Elmegreen, B. G. 2002, AJ, 123, 1454
\bibitem[Boselli et al.(1998)]{boselli98} Boselli, A., Lequeux, J.,
Sauvage, M., Boulade, O., Boulanger, F., Cesarsky, D., Dupraz, C.,
Madden, S., Viallefond, F., \& Vigroux, L. 1998, A\&A, 335, 53
\bibitem[Boulanger et al.(1988)]{boulanger88} Boulanger, F., Beichman, C.,
Desert, F. X., Helou, G., Perault, M., \& Ryter, C. 1988, ApJ, 332, 328
\bibitem[Calzetti et al.(2005)]{calzetti05} Calzetti, D., et al. 2005, ApJ, 
633, 871
\bibitem[Cannon et al.(2005)]{cannon05} Cannon, J. M., et al. 2005, ApJ, 630, L37
\bibitem[Churchwell et al.(2004)]{churchwell} Churchwell, E., et al. 2004, ApJS, 154, 322
\bibitem[Dale et al.(2005)]{dale05} Dale, D. A., et al. 2005, ApJ, 633, 857
\bibitem[de Blok \& Walter(2000)]{n682200} de Blok, W. J. G., \& Walter, F.
2000, ApJ, 537, L95
\bibitem[de Vaucouleurs et al.(1991)]{rc3} de Vaucouleurs, G., 
de Vaucouleurs, A., Corwin, H., Buta, R.,
Paturel, G., \& Fouqu\'e, P. 1991, Third Reference Catalogue of Bright
Galaxies (New York, Springer-Verlag) 
\bibitem[Draine(2005)]{draine05} Draine, B. T. 2005, in The Spitzer Science Center
2005 Conference: Infrared Diagnostics of Galaxy Evolution, in preparation
\bibitem[Engelbracht et al.(2005)]{engelbracht05} Engelbracht, C. W., 
Gordon, K. D., Rieke, G. H., Werner, M. W., Dale, D. A., \& Latter, W. B.
2005, ApJ, 628, L29
\bibitem[Fazio et al.(1982)]{fazio04} Fazio, G. G., et al. 2004, ApJS, 154, 10
\bibitem[F\"orster Schreiber et al.(2004)]{forster04}
F\"orster Schreiber, N. M., Roussel, H., Sauvage, M., \&
Charmandaris, V. 2004, A\&A, 419, 501
\bibitem[Galliano et al.(2003)]{galliano03} Galliano, F., Madden, S. C., 
Jones, A. P., Wilson, C. D., Bernard, J.-P., \& Le Peintre, F. 2003, A\&A, 407, 159
\bibitem[Hodge(1971)]{hodge71} Hodge, P. W. 1971, ARA\&A, 9, 35
\bibitem[Hogg et al.(2005)]{hogg05} Hogg, D. W., et al. 2005, ApJ, 624, 162
\bibitem[Houck et al.(2004)]{houck04} Houck, J. R., et al. 2004, ApJS, 154, 211
\bibitem[Hunter \& Elmegreen(2004)]{globalsfr} Hunter, D. A., \& 
Elmegreen, B. G. 2004, AJ, 128, 2170
\bibitem[Hunter \& Elmegreen(2006)]{ubvjhk} Hunter, D. A., \& 
Elmegreen, B. G. 2006, ApJS, 162, 49
\bibitem[Hunter et al.(2001)]{hunter01b} Hunter, D. A., \et\ 2001, ApJ, 553, 121
\bibitem[Hunter \& Gallagher(1992)]{n4449fil92} Hunter, D. A., \& Gallagher, 
J. S., III 1992, ApJ, 391, L9
\bibitem[Hunter \& Gallagher(1997)]{n4449fil97} Hunter, D. A., \& Gallagher, 
J. S., III 1997, 475, 65
\bibitem[Hunter et al.(1989)]{iras89} Hunter, D. A., Gallagher, J. S., III, Rice, 
W. L., \& Gillett, F. C. 1989, ApJ, 336, 152
\bibitem[Hunter et al.(1986)]{iras86} Hunter, D. A., Gillett, F., Gallagher, J. S.,
III, Rice, W. L., and Low, F. 1986, ApJ, 303, 171
\bibitem[Hunter \& Hoffman(1999)]{abund99} Hunter, D. A., \& Hoffman, L.
1999, AJ, 117, 2789
\bibitem[Hunter et al.(1999)]{n4449} Hunter, D. A., van Woerden, H., \& 
Gallagher, J. S. 1999, AJ, 118, 2184
\bibitem[Kennicutt et al.(2003)]{sings} Kennicutt, R. C., Jr., et al. 2003, 
PASP, 115, 928
\bibitem[Kennicutt \& Skillman(2001)]{kennicutt01} 
Kennicutt, R. C., \& Skillman, E. D. 2001, AJ, 121, 14614
\bibitem[Kniazev et al.(2000)]{kniazev00} Kniazev, A. Y., et al. 
2000, A\&A, 357, 101
\bibitem[Lada \& Lada(2003)]{lada03} Lada, C. J., \& Lada, E. A. 2003, ARA\&A, 41 57
\bibitem[Lee \& Skillman(2004)]{lee04} Lee, H., \& Skillman, E. D.
2004, ApJ, 614, 698
\bibitem[L\'eger \& Puget(1984)]{leger84} L\'eger, A., \& Puget, J. L. 
1984, A\&A, 137, L5
\bibitem[Li \& Draine(2002)]{li02} Li, A., \& Draine, B. T. 2002, ApJ, 572, 232
\bibitem[Lu et al.(2003)]{lu03} Lu, N., et al. 2003, ApJ, 588, 199
\bibitem[Madden(2000)]{madden00} Madden, S. C. 2000, New Astr Rev, 44, 249
\bibitem[Marston et al.(2004)]{marston} Marston, A. P., et al. 2004, ApJS, 154, 333
\bibitem[Massey et al.(1992)]{massey92}
Massey, P., Armandroff, T. E., \& Conti, P. S. 1992, AJ, 103, 1159
\bibitem[]{1421} Mo, H. J., Mao, S., \& White, S. D. M. 1998, MNRAS, 295, 319 
\bibitem[Nikolaev \& Weinberg(2000)]{lmc00} Nikolaev, S., \& Weinberg, M. D.
2000, ApJ, 542, 804
\bibitem[Pahre et al.(2004)]{pahre04} Pahre, M. A., Ashby, M. L. N., Fazio, 
G. G., \& Willner, S. P. 2004, ApJS, 154, 235
\bibitem[Pilyugin(2001)]{pilyugin01} Pilyugin, L. S. 2001, A\&A, 374, 412
\bibitem[Reach et al.(2005)]{reach} Reach, W. T., et al. 2005, PASP, 117, 978
\bibitem[Regan et al.(2004)]{regan04} Regan, M. W., et al. 2004, ApJS, 154, 204
\bibitem[Rho et al.(2006)]{rho} Rho, J., Reach, W. T., Lefloch, B., \& Fazio, G.
2006, ApJ, in press
\bibitem[Richer \& McCall(1995)]{richer95} Richer, M. G., \& McCall, M. L.
1995, ApJ, 445, 642
\bibitem[Rosenberg et al.(2006)]{rosenberg06} Rosenberg, J. L., Ashby, M. L. 
N., Salzer, J. J., \& Huang, J.-S. 2006, ApJ, 636, 742
\bibitem[Roussel et al.(2001)]{roussel01} Roussel, H., Gauvage, M., Vigroux, L.,
\& Bosma, A. 2001, A\&A, 372, 427
\bibitem[Skillman et al.(1997)]{skillman97} 
Skillman, E. D., Bomans, D. J., \& Kobulnicky, H. A. 
1997, ApJ, 474, 205
\bibitem[Spoon(2003)]{spoon03} Spoon, H. W. W. 2003, Ph.D. Thesis,
Rijksuniversiteit Groningen
\bibitem[Sturm et al.(2000)]{sturm00} Sturm, E., Lutz, D., Tran, D.,
Feuchtgruber, H., Genzel, R., Kunze, D., Moorwood, A. F. M., \&
Thornley, M. D. 2000, A\&A, 358, 481
\bibitem[Tacconi-Garman et al.(2005)]{tacconi05} Tacconi-Garman, L. E., 
et al. 2005, A\&A, 432, 91
\bibitem[Talent(1980)]{talent80} Talent, D. 1980, Ph.D. Thesis, Rice University
\bibitem[Thuan et al.(1999)]{thuan99} Thuan, T. X., Sauvage, M., \& Madden, S.
1999, ApJ, 516, 783
\bibitem[Uchida et al.(1998)]{uchida} Uchida, K. I., Sellgren, K., \& Werner, M.
1998, ApJ, 493, 109
\bibitem[Verstraete et al.(1996)]{verstraete96} 
Verstraete, L., Puget, J. L., Falgarone, E., Drapatz, S., Wright, C. M., 
\& Timmermann, R. 1996, A\&A, 315, L337
\bibitem[Wilcots \& Miller(1998)]{ic1098} Wilcots, E. M., \& Miller, B. W.
1998, AJ, 116, 2363
\bibitem[Willner et a.(2004)]{willner04} Willner, S. P., et al.\ 2004,
ApJS, 154, 222

\end{thebibliography}
\end{document}